\documentclass[aps,pra,twocolumn,showpacs,superscriptaddress]{revtex4}

\usepackage{cmap}
\usepackage{bm,amsmath,amssymb,latexsym,mathrsfs,graphicx,enumerate}
\usepackage[mathcal]{euscript}
\usepackage{hyperref}
\usepackage{epsfig}

\usepackage{amsmath}
\usepackage{graphics}
\usepackage{epstopdf}

\usepackage{caption}
\DeclareGraphicsExtensions{.bmp,.png,.jpg}

\begin{document}

\title[Nonlinear Modes of the Symmetric Hyperbolic Waveguide]
{Nonlinear TM Modes of the Symmetric Hyperbolic Slab Waveguide}

\author{Ekaterina I. Lyashko}
\affiliation{\normalsize \noindent Department of General and Applied Physics,
Moscow Institute of Physics and Technology, Dolgoprudny, Moscow
region, 141700 Russia \\
E-mails: ostroukhova.ei@gmail.com }
\author{Andrei I. Maimistov}
\affiliation{\normalsize \noindent Department of Solid State Physics
and Nanostructures, National
Nuclear Research University,\\
Moscow Engineering Physics Institute, Moscow, 115409 }
\affiliation{\normalsize \noindent Department of General Physics,
Moscow Institute of Physics and Technology, Dolgoprudny, Moscow region, 141700 Russia \\
E-mails: aimaimistov@gmail.com }

\date{\today}

\begin{abstract}
Guided wave modes in a symmetric slab waveguide formed by an
isotropic dielectric layer with cubic nonlinear response placed in
the hyperbolic surrounding medium are investigated theoretically.
Optical axis of the hyperbolic medium is normal to the slab plane.
If dielectric permittivity of the waveguide core is more than
extraordinary permittivity of the hyperbolic medium each TM mode has
two cut-off frequencies. The dispersion relations for these modes
are found and numerically solved for cases of focusing and
defocussing medium of the core. Number of possible modes at given
frequency depends on the applied field intensity.
It is shown that zero values of the
TM modes propagation constants are possible in the waveguide.
Moreover in the self-defocussing case they can be obtained by
increasing field intensity. The dependence of a propagation constant
and width of the mode on radiation intensity is obtained and
analyzed.
\end{abstract}


%\keywords{nonlinear optics, guided waves, anisotropy, hyperbolic material}

%\submitto{\JOPT}

\maketitle

%\ioptwocol

%42.65.Tg    Optical solitons; nonlinear guided waves (for solitons
%            in fibers, see 42.81.Dp)
%42.79.Gn    (Optical waveguides and couplers)
%42.70.Qs    Photonic bandgap materials (for photonic crystal lasers,
%see 42.55.Tv)
%42.81.Qb    (Fiber  waveguides, couplers, and arrays)

\section{Introduction}

Hyperbolic medium can be defined as strongly anisotropic uniaxial
medium which principal components of dielectric permittivity or
magnetic permeability tensors have opposite signs
\cite{Narim:06,Noginov:09}. In case of nonmagnetic hyperbolic
material an extraordinary wave propagating through this medium has a
hyperbolic dispersion. This lead to a number of new optical
%(electrodynamic?)
phenomena \cite{Poddubny:12,Poddubny:13,Ferrari:14,Benedict:13,
Zapata:13,XNi:11}. A review is presented in
\cite{Drachev:13,Shekhar:14}.

Photonic or plasmonic guiding devices, such as waveguides, are
widely used in different communication networks or information
processing systems. A numerous guiding structures formed from
hyperbolic medium as component were investigated already. Among them
are plasmonic waveguides with hyperbolic cladding
\cite{Kildishev:15a,Kildishev:15b}, photonic waveguides with
hyperbolic core \cite{Huang:08,Zhu:15} or surroundings
\cite{Guo-ding Xu:08} demonstrating negative refractive
index, slow light, large mode index, etc. All these works are
limited by linear media response to the applied radiation.

In our previous work \cite{Lyashko:15} a slab waveguide with
isotropic dielectric core in the hyperbolic host was considered. In
this structure much of the guided wave energy is concentrated in the
non-dissipative core, whereas only evanescent waves are in the
hyperbolic medium. The optical axis of hyperbolic medium is aligned
with normal vector to the wave propagation direction. In this
geometry the transverse magnetic polarization modes is extraordinary
waves in the hyperbolic cladding. This leads to the new phenomenon:
each TM guiding wave could be characterized by the two cut-off
frequencies. Thus, each TM mode is held by the waveguide only in
certain frequency range or the mode exists only in certain interval
of core thickness. It is conventional only one cut-off frequency
exists and the number of the guided waves is always increasing with
the frequency or the core thickness.

\begin{figure}[ht]
        \centering
        \includegraphics[width=0.75\linewidth]{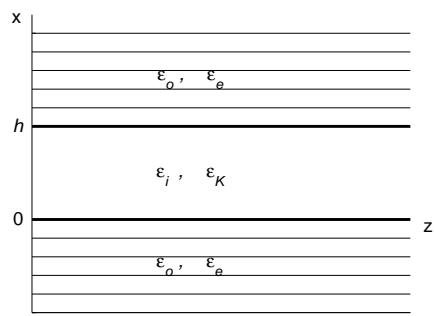}
    \caption{Scheme of hyperbolic slab waveguide}
    \label{Fig1:scheme}
\end{figure}

Nonlinear guided wave modes in an \emph{asymmetric slab waveguide}
formed by an isotropic dielectric layer placed on a linear or
nonlinear substrate and covered by a hyperbolic material were
investigated in \cite{LM:16}. In this case the additional solutions
for guided waves arises. These modes corresponds with situation,
where the peak of electric field is localized in the nonlinear
substrate. These modes are absent in the linear waveguide. To excite
these modes the power must exceed certain threshold value.  The
cut-off frequencies for each mode are determined by the dielectric
permittivities and the field intensity.

In present paper the nonlinear guided wave modes in an
\emph{symmetric slab waveguide} formed by an isotropic nonlinear
dielectric layer dipped into a hyperbolic material will be
considered.

In the next step we will consider a cubic nonlinear response of the
core medium and its influence on the waveguide dispersion
characteristics. The field distributions and dispersion relations,
connecting mode propagation constant and radiation frequency, are
analytically obtained and analyzed. Cases of self-focusing and
self-defocusing Kerr medium of the core are considered. In all cases
the two cutoff frequencies phenomenon also exist. In the case of
focusing core medium mode effective refractive index (or propagation
constant) increases with field intensity growth. In the de-focusing
case propagation constant monotonically decreases with intensity
growth, achieving zero value. This situation equals to zero energy
flux in the mode propagation direction or "stopped light". Also it
is possible to control a number of propagating modes in the
hyperbolic waveguide varying the field intensity. There is another
phenomenon in case of self-defocusing core in the hyperbolic
environment, namely, the mode width narrows with field intensity
growth.

\section{Constituent Equations}

\noindent It is assumed that waveguide under consideration is formed
by nonmagnetic media without any dissipation. The substrate and the
covering layer are uniaxial hyperbolic media with an optical axis
directed perpendicularly to the media interfaces. An electromagnetic
properties of the hyperbolic media are described by principle
components of the dielectric tensor: $\varepsilon_o$ and
$\varepsilon_e$. The core is a cubic nonlinear (Kerr) isotropic
dielectric characterized by linear dielectric permittivity
$\varepsilon_i$ and Kerr constant $\varepsilon_K $. The coordinate
axes are determined as follows. $\textit{OX}$ axis is directed along
with the normal vector to the interfaces. $\textit{OY}$ and
$\textit{OZ}$ axes are parallel to the core interfaces. It is
assumed that guided waves propagate along the $\textit{OZ}$ axis. In
such planar geometry the electromagnetic fields are independent on
the $\textit{y}$ coordinate.

To describe an electromagnetic wave propagation in the general case
following system can be obtained from the wave equations
\cite{Lyashko:15,LM:16}
\begin{eqnarray}
&& \left(\frac{\varepsilon_e(x)}{\varepsilon_o(x)}\frac{\partial^2}{\partial
    x^2} +\frac{\partial^2}{\partial z^2}
+k_0^2\varepsilon_e(x)\right)E_x +4\pi k_0^2P_{nl,x} =0, \nonumber\\
&& \left(\frac{\partial^2}{\partial x^2} +\frac{\partial^2}{\partial z^2}
+k_0^2\varepsilon_o(x)\right)E_y +4\pi k_0^2P_{nl,y}
=0,\label{eq:we} \\
&& \left(\frac{\partial^2}{\partial x^2} +\frac{\partial^2}{\partial
z^2} +k_0^2\varepsilon_o(x)\right)E_z
+\frac{\Delta\varepsilon(x)}{\varepsilon_o(x)}
\frac{\partial^2}{\partial x\partial z}E_x + \nonumber\\
&& \qquad \qquad \qquad \qquad \qquad  + 4\pi k_0^2P_{nl,z} =0.
\nonumber
\end{eqnarray}

If planar waveguide with core width $h$ is considered, linear
principal dielectric constants can be presented as piecewise
functions
\begin{equation}\label{eq:epsl}
\varepsilon_{o,e}(x)=\left\{
\begin{array}{ccc}
\varepsilon_{o,e} & ~~ x<0, \\
\varepsilon_i & ~~ 0 \leq x \leq h, \\
\varepsilon_{o,e} & ~~ x>h, \\
\end{array}\right.
\end{equation}
In the planar geometry under consideration (Fig. \ref{Fig1:scheme})
transverse electric (TE) and transverse magnetic (TM) waves can be
analyzed independently. The present work is dedicated to TM
polarization case. TM wave is an extraordinary wave and is defined
by following field components $\mathbf{E}=(E_x, 0, E_z)$ and
$\mathbf{H}=(0, H_y, 0)$. These components are connected with each
other
\begin{eqnarray}
     H_y &=&\frac{i}{k_0}\left(\frac{\partial E_z}{\partial x}-\frac{\partial E_x}{\partial z}\right), \nonumber \\
     \frac{\partial H_y}{\partial z} &=& ~~~ik_0\varepsilon_e(x)E_x+i 4\pi k_0 P_{nl,
    x}, \label{eq:con}\\
    \frac{\partial H_y}{\partial x} &=& -ik_0 \varepsilon_o(x)E_z
    -i4\pi k_0 P_{nl, z}. \nonumber
\end{eqnarray}

It is assumed that no harmonic generation processes available in the
waveguide core. Nonlinear polarization for TM wave in case of
isotropic core medium can be presented as
\begin{eqnarray}
\nonumber
P_x^{nl}= (\gamma_{xx}|E_x|^2+\gamma_{xz}|E_z|^2)E_x, \\
\nonumber
P_z^{nl}= (\gamma_{zz}|E_z|^2+\gamma_{zx}|E_x|^2)E_z,
\end{eqnarray}
where parameters $\gamma_{ij}$ depend on forth-rank susceptibility
tensor components $\chi_{mnkl}$. In further analysis for simplicity
we will consider uniaxial approximation
\cite{Mihala:Fedy:83,Agranowich:86} $\gamma_{zz} = \varepsilon_K /
4\pi$, $\gamma_{xz} = \gamma_{zx} = \gamma_{xx} = 0$. Thus, an
expression for nonlinear polarization will take a form:
\begin{equation} \label{eq:nlp}
\textbf{P}^{nl} = \frac{\varepsilon_K}{4\pi} |E_z|^2 E_z \textbf{e}_z.
\end{equation}

\section{Electromagnetic Field Distribution}

\noindent The considered waveguide is homogeneous in $Z$ direction.
So the guiding wave fields can be presented as
$\mathbf{E}=(\tilde{E}_x, 0, \tilde{E}_z)e^{i\beta z}$ and
$\mathbf{H}=(0, \tilde{H}_y,0)e^{i\beta z}$, where $\beta$ is
propagation constant. In this assumption one can obtain from
(\ref{eq:we}) the following system for the $\tilde{E}_z$ component
\begin{eqnarray}
x<0   &~& \frac{\partial^2}{\partial x^2}\tilde{E}_z+\frac{\varepsilon_o}{\varepsilon_e}(k_0^2\varepsilon_e - \beta^2)\tilde{E}_z =0,   \nonumber\\
0\leq x\leq h &~& \frac{\partial^2}{\partial x^2}\tilde{E}_z+(k_0^2\varepsilon_i - \beta^2)\tilde{E}_z +k_0^2 \varepsilon_K \tilde{E}_z^3=0, \label{eq:Ez:1}\\
x>h   &~& \frac{\partial^2}{\partial
x^2}\tilde{E}_z+\frac{\varepsilon_o}{\varepsilon_e}\left(k_0^2\varepsilon_e
- \beta^2\right)\tilde{E}_z =0,  \nonumber
\end{eqnarray}
It is suitable to introduce new parameters $q^2 = \varepsilon_o /
\varepsilon_e(\beta^2 - k_0^2\varepsilon_e)$, $p^2 = k_0^2
\varepsilon_i - \beta^2$. To consider guided modes of the waveguide
the condition $q^2 > 0$ must be satisfied. In the case of a
hyperbolic material this inequality is satisfied only when
$\varepsilon_o < 0$ and $\varepsilon_e > 0$.

There is no restriction for the $p^2$ sign. The case of $p^2 < 0$
corresponds to pair of two surface waves propagating along the
waveguide interfaces. The case $p^2 > 0$ corresponds to guided modes
regime of the waveguide. In further analysis the case $p^2 >0$ will
be considered.

The first and the last equations of the system (\ref{eq:Ez:1}) can
be simply integrated
\begin{equation} \label{eq:E1E3}
\tilde{E}_z^{(1)} = A_0 e^{qx}, \quad \tilde{E}_z^{(3)} = A_h
e^{-qx},
\end{equation}
The upper marks 1 and 3 are referred to substrate and covering layer
respectively.

The second equation of (\ref{eq:Ez:1}) can be integrated in terms of
Jacobi elliptic functions \cite{Batm:Erd:67}. The first step is to
present (\ref{eq:Ez:1}) as an integral of motion
\begin{equation}
\label{eq:int:1} \left(\frac{\partial}{\partial x} \tilde{E}_z
\right)^2 + p^2 \tilde{E}_z^2 + \frac{1}{2} k_0^2 \varepsilon_K
\tilde{E}_z^4 = G,
\end{equation}
where $G$ is an integration constant.

In further derivation we will concentrate on case of the
self-defocusing core medium, $\varepsilon_K < 0$. Case of
self-focusing medium could be described in similar way.

By introducing a new variable $v = \sqrt{\alpha} \tilde{E}_z$, where
$\alpha = \sqrt{k_0^2 |\varepsilon_K|/2}$ is a new parameter of
nonlinearity, one can obtain from the (\ref{eq:int:1})
\begin{equation} \label{eq:J1}
\left( \frac{\partial v}{\partial x} \right)^2 = \alpha (v^2 -
\bar{v}_1^2)(v^2 - \bar{v}_2^2).
\end{equation}
Here $\bar{v}_1^2$, $\bar{v}_2^2$ are
\begin{equation}%\label{eq:roots}
    \bar{v}_1^2 = \frac{p^2}{2 \alpha} - \sqrt{ \left( \frac{p^2}{2
\alpha} \right)^2 - G},~~~ \bar{v}_2^2 = \frac{p^2}{2 \alpha} +
\sqrt{ \left( \frac{p^2}{2 \alpha} \right)^2 - G}. \nonumber
\end{equation}

By the use of $w(x) = v(x)/\bar{v}_1$ the equation (\ref{eq:J1}) can
be presented as
\begin{equation} \label{eq:dif:w}
\left( \frac{\partial w}{\partial x} \right)^2 = \alpha \bar{v}_2^2
(1-w^2)(1 - \xi^2w^2),
\end{equation}
where $ \xi^2 = \bar{v}_1^2/\bar{v}_2^2$, $0 \leq \xi^2 \leq 1$, and
$\xi$ monotonically increases with $G$. Hence parameter $\xi$ can be
considered as a new integration constant. Parameters in
(\ref{eq:J1}) can be presented in terms of $\xi$
$$
\bar{v}_1^2 = \frac{p^2}{\alpha} \frac{\xi^2}{1+ \xi^2}, ~~~~
\bar{v}_2^2 = \frac{p^2}{\alpha} \frac{1}{1+\xi^2}.
$$

By integrating (\ref{eq:dif:w}) one obtains the expression
\begin{equation}
\label{eq:J2} \pm \frac{p (x-x_0)}{\sqrt{1+\xi^2}} = \int_{0}^{w}
\frac{d \tilde{w}}{\sqrt{(1-\tilde{w}^2)(1-\xi^2 \tilde{w}^2)}},
\end{equation}
where $x_0$ is integration constant. Integral in the right side of
the equation is the incomplete elliptic integral of the first genus
\cite{Batm:Erd:67}. Hence $w$ can be expressed in terms of the
Jacobi elliptic sinus
$$ w (x) = \pm \mathrm{sn} \left(\frac{p
(x-x_0)}{\sqrt{1+\xi^2}}, ~ \xi \right).
$$
Z-component of electric field in the core of waveguide is defined as
$$
 \tilde{E}_z^{(2)} = \frac{\bar{v}_1}{\sqrt{\alpha}} w(x) =
\pm \frac{p}{\alpha} \frac{\xi}{\sqrt{1+\xi^2}} ~ \mathrm{sn} \left(
\frac{p(x - x_0)}{\sqrt{1+\xi^2}} ; ~ \xi \right).
$$
It is suitable to introduce a new variable
\begin{equation} \label{eq:Am}
A_{m} = \frac{p}{\alpha} \frac{\xi}{\sqrt{1+\xi^2}},
\end{equation}
which is the maximum value of field component $E_z$.

As is known, tangential components of electric and magnetic fields
are continuous functions on the boundaries between media. Taking
into account the equations (\ref{eq:E1E3}) for electric fields in
the substrate and in the cover, the following definitions can be
achieved
\begin{eqnarray}
\nonumber  %to remove numbering (before each equation)
x<0   &~& \tilde{E}^{(1)}_z(x) = \pm A_{m} ~
 \mathrm{sn} \left( \frac{- px_0}{\sqrt{1+\xi^2}} ; ~ \xi \right) e^{q x}, \\
0\leq x\leq h &~& \tilde{E}^{(2)}_z(x) = \pm A_{m} ~
\mathrm{sn} \left( \frac{p(x - x_0)}{\sqrt{1+\xi^2}} ; ~ \xi \right),  \label{eq:DEz}\\
\nonumber x>h   &~& \tilde{E}^{(3)}_z(x) = \pm A_{m} ~ \mathrm{sn}
\left( \frac{p(h - x_0)}{\sqrt{1+\xi^2}} ; ~ \xi \right) e^{q(h-x)},
\end{eqnarray}
so $\tilde{E}_z^{(1)}(0) = \tilde{E}_z^{(2)}(0)$ and
$\tilde{E}_z^{(2)}(h) = \tilde{E}_z^{(3)}(h)$. The upper marks 1, 2
and 3 are referred to substrate, core and covering layer.

In the case of self-focusing core medium the Z-components of
electric field distribution can be obtained by the use of
substitution $\xi = i \kappa$, where $0 \leq \kappa \leq 1$.

Components of $\tilde{H}_y$ and $\tilde{E}_x$ can be easy expressed
in terms of $\tilde{E}_z$ using the following formulas:
\begin{eqnarray}
&& \tilde{H}_y = \frac{i k_0 \varepsilon_e(x)}{k_0^2
\varepsilon_e(x) - \beta^2} \frac{\partial}{\partial x}\tilde{E}_z, \nonumber\\
&& \label{eq:Hy} \\
&& \tilde{E}_x = \frac{i \beta}{k_0^2 \varepsilon_e(x) - \beta^2}
\frac{\partial}{\partial x}\tilde{E}_z, \nonumber
\end{eqnarray}
that follow from the system (\ref{eq:con}) and assumption
(\ref{eq:nlp}).

It is important to understand how the system (\ref{eq:DEz}) can be
transformed to the linear case. Let us consider the case when
$\varepsilon_K$ is negligibly small. Thus parameter $\xi$ will be:
$$
\xi^2 = \frac{p^2 - \sqrt{p^4 - G (2\alpha)^2}}{p^2 +
\sqrt{p^4 - G (2\alpha)^2}} ~ \to 0 ~~ (\alpha \to 0).
$$
Then
$$
 \mathrm{sn} ( p(x - x_0) ; ~ 0) ~ \to \sin(p(x - x_0)).
$$ So maximum value of $\tilde{E}_z^{(1,2,3)} = A_{m}$
will be
\begin{equation}
A_{m}^2 = \frac{p^2}{\alpha^2} \frac{\xi^2}{1+\xi^2} ~ \to ~
\frac{G}{p^2}, \label{eq:Am:2}
\end{equation}
where the definition of $\xi$ and L'H\^{o}pital's rule were used.
Hence, $G$ is connected with maximum field intensity in the linear case.

\section{Dispersion Relation}

\noindent Taking into account the continuity conditions for
$\tilde{H}_y$ at the core boundaries ($x = 0$, $x = h$), equations
(\ref{eq:Hy}) and system (\ref{eq:DEz}) the following relations can
be achieved
\begin{eqnarray}
&& \qquad \qquad \qquad - \frac{\varepsilon_o}{\varepsilon_i}
\frac{p}{q} \sqrt{1+\xi^2} =
\nonumber \\
&&\frac{\mathrm{cn}\left(-px_0/\sqrt{1+\xi^2} ; \xi \right)
\mathrm{dn}\left(-px_0/\sqrt{1+\xi^2} ; ~ \xi
\right)}{\mathrm{sn}\left( -px_0/\sqrt{1+\xi^2} ; \xi \right)},
\nonumber\\
&&\label{eq:disp}\\
&&\qquad \qquad \qquad \frac{\varepsilon_o}{\varepsilon_i}
\frac{p}{q} \sqrt{1+\xi^2}
=\nonumber\\
&& \frac{\mathrm{cn}\left(p(h-x_0)/ \sqrt{1+\xi^2} ; \xi \right)
\mathrm{dn}\left(p(h-x_0)/ \sqrt{1+\xi^2} ; ~ \xi
\right)}{\mathrm{sn}\left(p(h-x_0)/ \sqrt{1+\xi^2} ; \xi \right)}.
\nonumber
\end{eqnarray}
Comparison of the equations of this system results in the relation
$x_0 = h/2 + NT\sqrt{1+\xi^2}/p$, where $T$ is a period of elliptic
function $\mathrm{sn}(x, \xi)$ and $N$ is an integer constant.
Periods of the elliptic functions $\mathrm{sn}(x, \xi)$ and
$\mathrm{cn}(x, \xi)$ are the same. Period of $\mathrm{dn}(x, \xi)$
is equal to $T/2$. Thus the system (\ref{eq:disp}) can be reduced to
one equation
\begin{eqnarray}
&&\qquad \qquad \qquad \frac{\varepsilon_o}{\varepsilon_i}
\frac{p}{q} \sqrt{1+\xi^2} =
\label{eq:disp2}\\
&& =\frac{\mathrm{cn}\left(ph/2 \sqrt{1+\xi^2} ; ~ \xi \right)
\mathrm{dn}\left(ph/2 \sqrt{1+\xi^2} ; ~ \xi
\right)}{\mathrm{sn}\left(ph/2 \sqrt{1+\xi^2} ; ~ \xi
\right)}.\nonumber
\end{eqnarray}
This equation connects propagation constant $\beta$ included in
parameters $p$ and $q$ with radiation frequency $\omega$ included in
the wave number $k_0$. Hence obtained equation is the dispersion
relation. The substitution $\xi = i\kappa$ results in the dispersion
relation for the self-focusing core medium case, $\varepsilon_K>0$.

Parameter $\xi$ accounts nonlinear property of the
waveguide core. As follows from (\ref{eq:Am}) its value increases
with intensity $A_m^2$.

In the linear case, $\xi = 0$, elliptic functions are replaced by
theirs trigonometrical analogous and relation (\ref{eq:disp2})
reduces to the dispersion equation obtained earlier \cite{LM:16}.

\begin{figure}[h!]
    \centering
    \includegraphics[width=.8\linewidth]{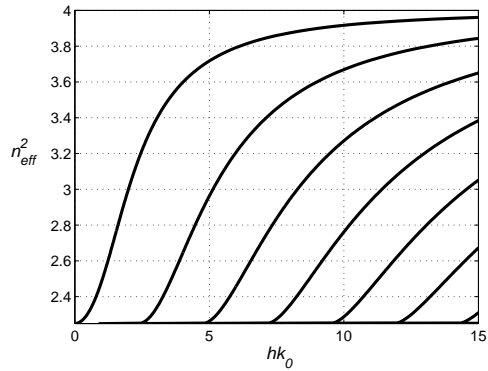}
    \caption{Dispersion curves for TM modes of the standard dielectric waveguide.
    Mode marks are: $m = 0, 1, 2, 3, 4, 5$.}
    \label{Fig2:ldisp}
\end{figure}

In further analysis we will consider the mode effective refractive
index $n_{eff} = \beta / k_0$ instead of mode propagation constant
$\beta$. In the previous section it was shown that propagation
parameters $q$ and $p$ must satisfy inequalities $q^2 > 0$, $p^2 >
0$. In the case of hyperbolic claddings with dielectric constants
$\varepsilon_o < 0$, $\varepsilon_e > 0$ these conditions set limits
on possible values of $n_{eff}$:
\begin{equation} \label{eq:nef:1}
    0 \leq n_{eff}^2 \leq \min{(\varepsilon_e, \varepsilon_i)},
\end{equation}
were value $n_{eff} = 0$ means zero value of wave vector projection
on the mode propagation direction ($Z$ axis), i.e. stopped wave
between the waveguide claddings. For comparison in the usual case of
waveguide with dielectric claddings possible values of $n_{eff}$ are
in the range
\begin{equation} \label{eq:nef:2}
    \varepsilon_{e} \leq n_{eff}^2 \leq \varepsilon_{i}.
\end{equation}

It is interesting to note that value $n_{eff} = \varepsilon_i$ or $p
= 0$ is a solution of dispersion relations both for hyperbolic and
standard dielectric waveguide cases for all values of frequency or
core width $h$. The elliptic functions presented in (\ref{eq:disp2})
are the periodic functions excepting the case $\xi=1$. Thus equation
(\ref{eq:disp2}) has a number of different solution brunches
$n_{eff}^2 (hk_0)$ that are dispersion characteristics for the
different waveguide modes. They are usually marked with integer mode
mark $m$. If the value $n_{eff}^2 = \varepsilon_i$ is possible it
will be at least one mutual point for all brunches. In
Fig.\ref{Fig2:ldisp} the dispersion curves for TM modes of a usual
linear dielectric slab waveguide with $\varepsilon_i = 4.0$,
$\varepsilon_e = \varepsilon_o = 2.25$ are presented. One can notice
that each TM mode curve have a starting point (cutoff frequency)
where $n_{eff}^2 = \varepsilon_e$ and then tends to the common value
$n_{eff}^2 = \varepsilon_i$ that is achieved at $hk_0 \to \infty$.
In the hyperbolic waveguide case at $\varepsilon_e < \varepsilon_i$,
the value $n_{eff}^2 = \varepsilon_i$ is not possible due to
(\ref{eq:nef:2}) and the behavior of dispersion curves is quite
different.

\begin{figure}[h!]
    \centering
    \begin{minipage}[ht]{.8\linewidth}
        \centering
        \includegraphics[width=\linewidth]{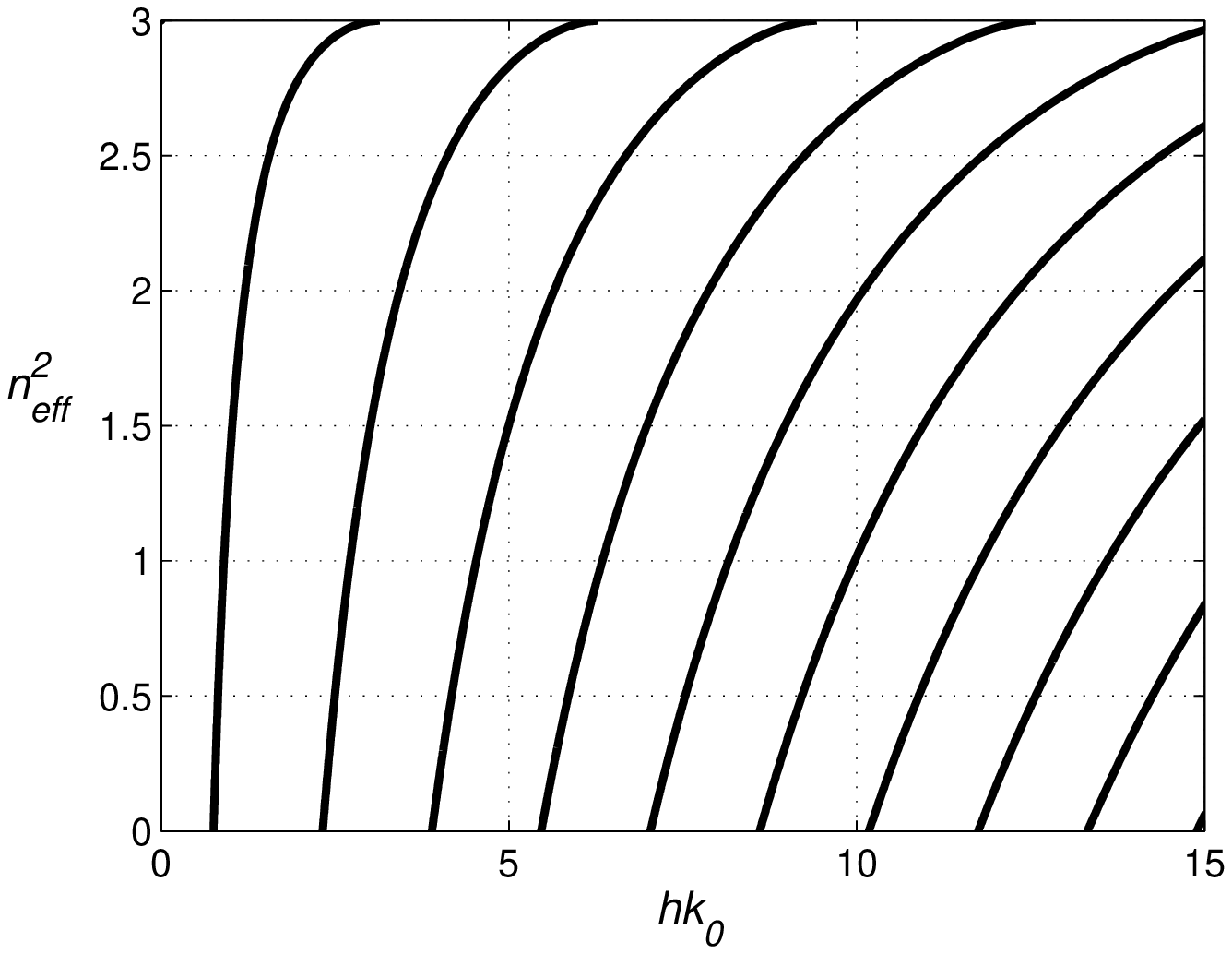}
        \\(\textit{a}) $\xi = 0$
    \end{minipage}
    \vfill
    \begin{minipage}[ht]{.8\linewidth}
        \centering
        \includegraphics[width=\linewidth]{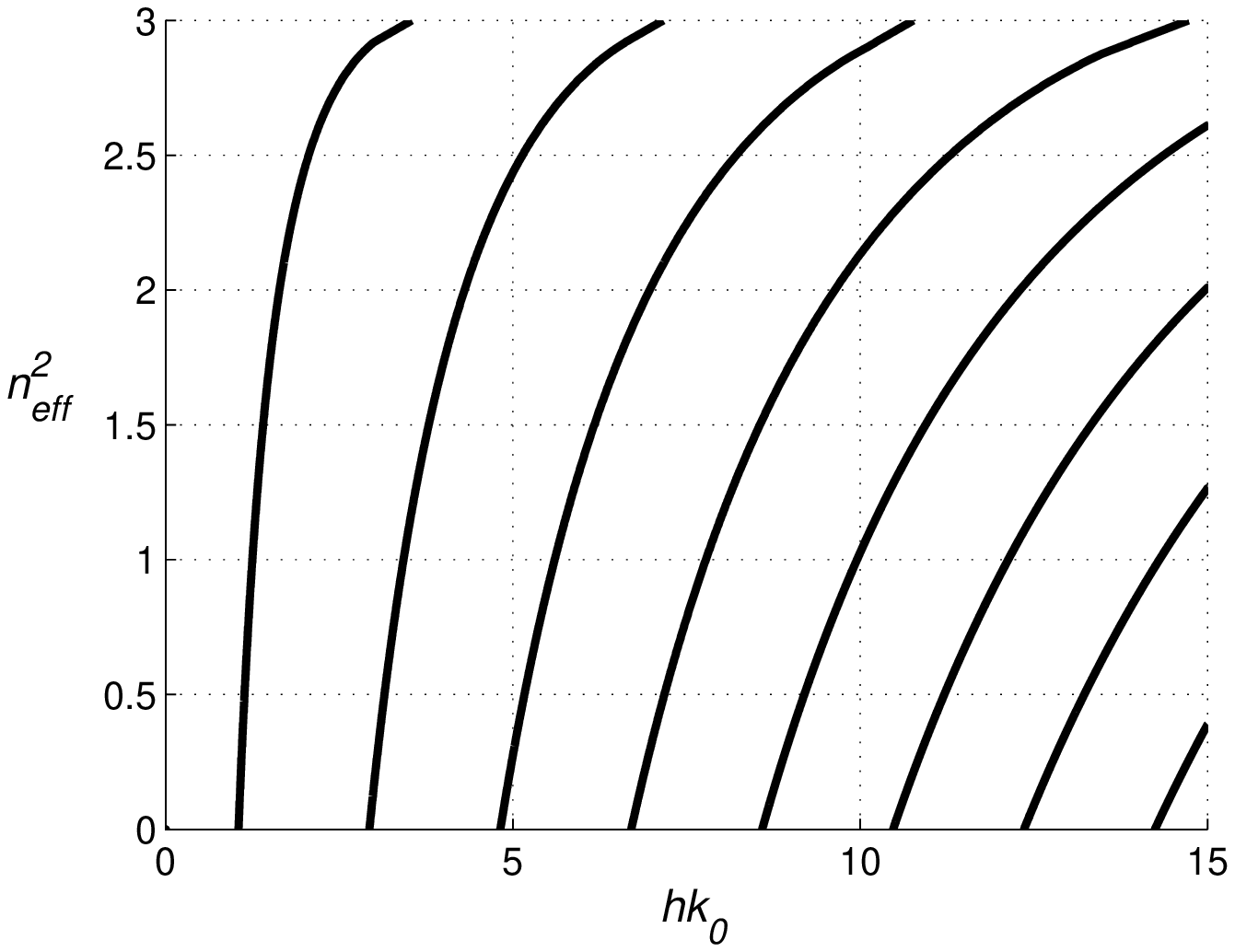}
        \\(\textit{b}) $\xi = 0.5$
    \end{minipage}
    \vfill
    \begin{minipage}[ht]{.8\linewidth}
        \centering
        \includegraphics[width=\linewidth]{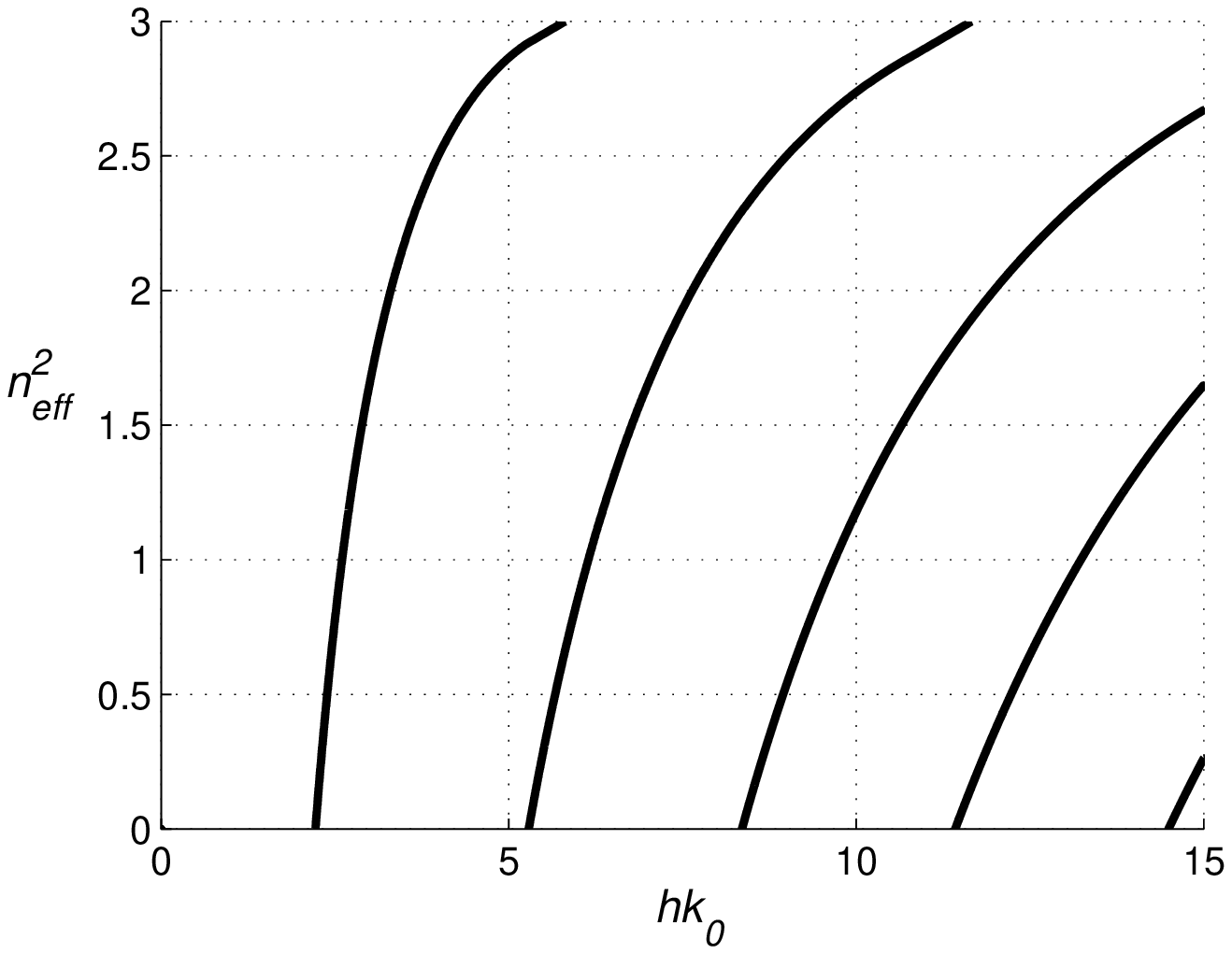}
        \\(\textit{c}) $\xi = 0.9$
    \end{minipage}
    \caption{Dispersion curves for TM modes in case of $\varepsilon_K < 0$, $\varepsilon_e < \varepsilon_i$.
        Mode marks are: (\textit{a}) $m$ = 1, 2,\ldots, 9, (\textit{b}) $m$ = 1, 2,\ldots, 8, (\textit{c}) $m$ = 1, 2,\ldots, 5} .
    \label{Fig3:disp:1}
\end{figure}

Functions $n_{eff}^2 (hk_0)$ that are numerical solutions of the
relation (\ref{eq:disp2}) are presented in Fig. \ref{Fig3:disp:1}
for different values of $\xi$. Linear dielectric permittivities of
the waveguide layers are $\varepsilon_i = 4.0$, $\varepsilon_e =
3.0$, $\varepsilon_o = -3.5$. In several works \cite{Guo-ding
Xu:08,Lyashko:15} it was shown that in the waveguide with hyperbolic
environment there is not a fundamental mode corresponding to $m=0$.
So in each case presented in Fig. \ref{Fig3:disp:1} mode index
starts from $m=1$ value.

From the pictures presented in Fig.\ref{Fig3:disp:1} it follows that
TM modes in each case have two cutoff frequencies. At
$n_{eff}^2(hk_0) = 0$ the mode appears in the waveguide, at
$n_{eff}^2(hk_0) = \varepsilon_e$ the mode leaves the waveguide.
That is not possible in the usual dielectric waveguides. An
additional cutoff frequency phenomenon found earlier in the linear
case \cite{Lyashko:15} is present in the case of waveguide with
nonlinear core too.

By comparing results presented on pictures of Fig.\ref{Fig3:disp:1}
(a), (b), and (c) one can notice that with nonlinear parameter $\xi$
growth the density of dispersion characteristics becomes lower and
each curve shifts to the greater values of normalized core width (or
frequency). This allows to make conclusion that with intensity
growth in the self-defocusing case an effective mode index $n_{eff}$
or propagation constant $\beta$ decreases and finally achieves zero
value (for constant frequency value). Then mode leaves the
waveguide. A case $\beta = 0$ equals to the stopped waveguide mode.

\begin{figure}[h!]
\centering
    \begin{minipage}[h]{.8\linewidth}
        \centering
        \includegraphics[width=\linewidth]{fig3a.eps}
        \\(\textit{a}) $\kappa = 0$
    \end{minipage}
    \vfill
    \begin{minipage}[h]{.8\linewidth}
        \centering
        \includegraphics[width=\linewidth]{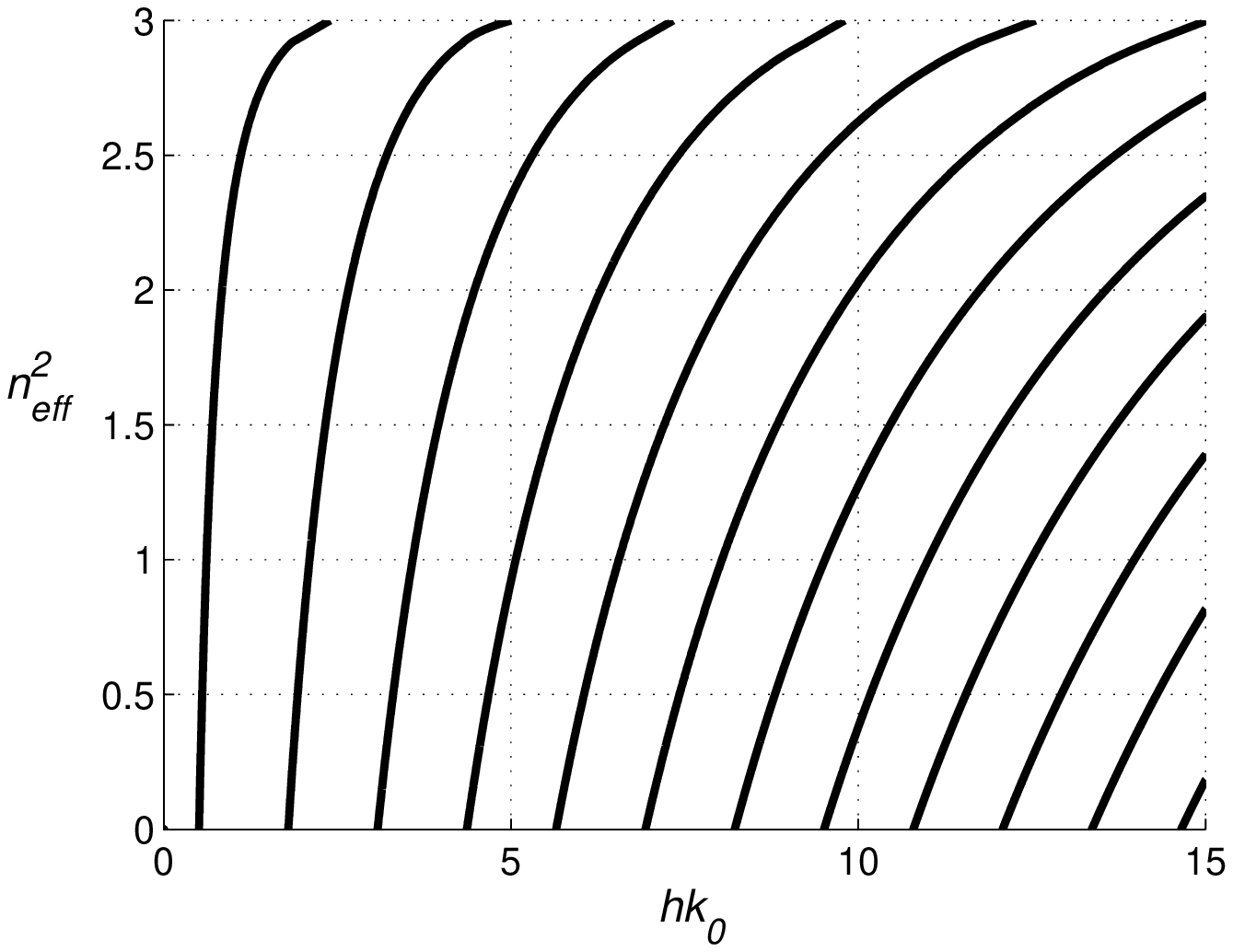}
        \\(\textit{b}) $\kappa = 0.5$
    \end{minipage}
    \vfill
    \begin{minipage}[h]{.8\linewidth}
        \centering
        \includegraphics[width=\linewidth]{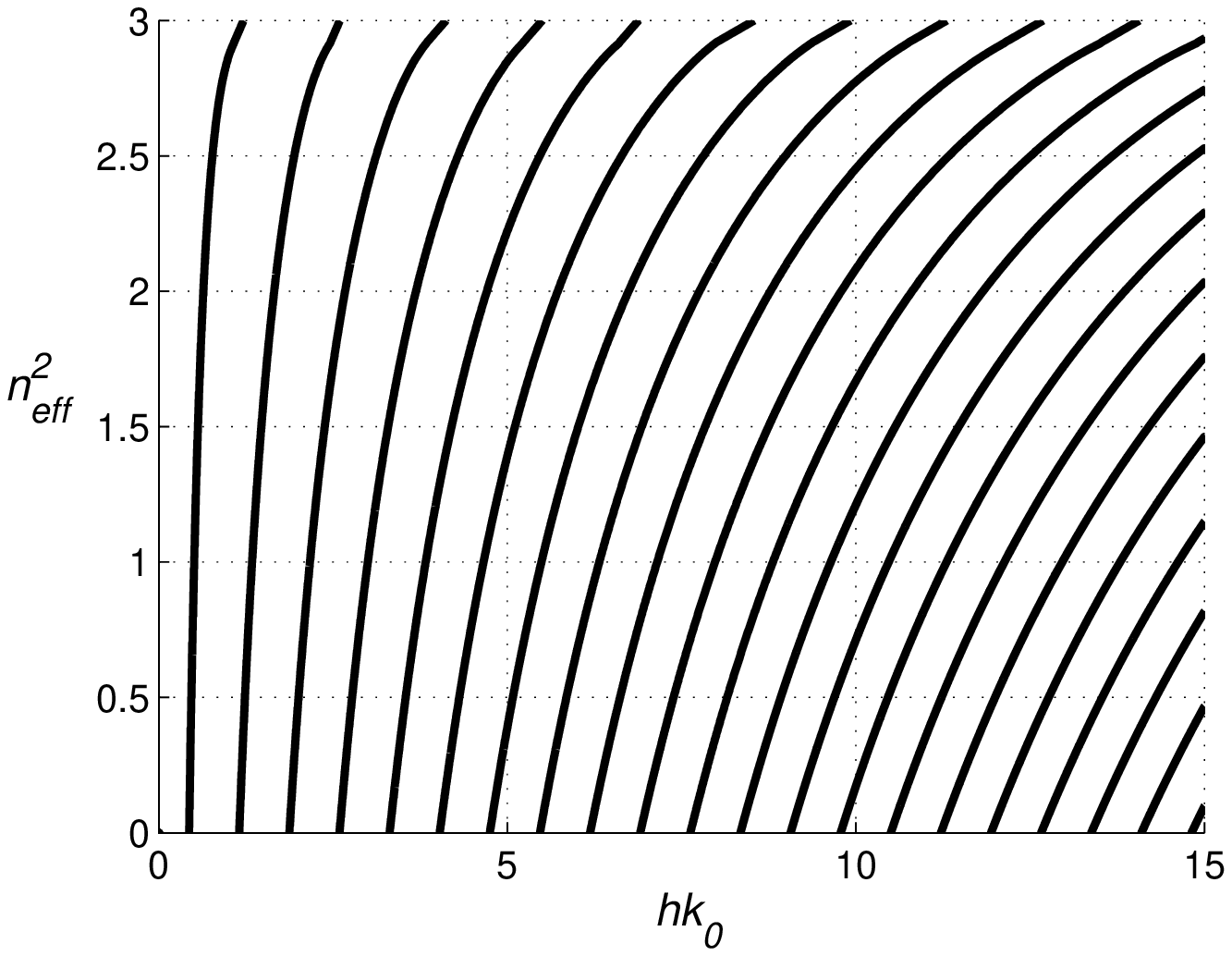}
        \\(\textit{c}) $\kappa = 0.85$
    \end{minipage}
    \caption{Dispersion curves for TM modes in case of $\varepsilon_K > 0$, $\varepsilon_e < \varepsilon_i$.
    Mode marks are: (\textit{a}) $m$ = 1, 2,\ldots, 9, (\textit{b}) $m$ = 1, 2,\ldots, 12, (\textit{c}) $m$ = 1, 2,\ldots, 21.}
    \label{Fig4:disp:2}
\end{figure}

In Fig. \ref{Fig4:disp:2} dispersion curves for TM waves in case of
$\varepsilon_K > 0$ for different values of parameter $\kappa$ are
presented. From the pictures follows that each TM mode also has two
cutoff frequencies. But in comparison with self-defocusing case the
dispersion curves density becomes higher with nonlinear parameter
growth. And each curve shifts to the lower values of normalized core
width or frequency. Thus in the focusing case a value of $n_{eff}^2$
increases with intensity at constant radiation frequency (or core
width) finally achieves $\varepsilon_e$ value. Then considered mode
leaves the waveguide.

Comparing the results obtained for cases $\varepsilon_K < 0$ and
$\varepsilon_K > 0$ the first one seems to be more interesting: the
number of possible solutions here decreases with intensity growth.
This fact could prevent mode propagation from the inter-modes
dispersion. And slowing light is possible.

In further parts of the paper we will analyze the case of slab
waveguide with self-defocusing medium of the core in more details.

\subsection{The case of $\xi = 1$}

At maximum possible field intensity a nonlinear parameter $\xi$ is
equal to value $\xi = 1$. In this case the elliptic functions in the
dispersion relation can be replaced by corresponding hyperbolic
functions \cite{Batm:Erd:67}. Thus equation (\ref{eq:disp2}) is
reduced to the form
\begin{eqnarray}
&&\tanh \left( \frac{hk_0}{2\sqrt{2}} \sqrt{\varepsilon_i -
n_{eff}^2} \right) \cosh^2 \left( \frac{hk_0}{2\sqrt{2}}
\sqrt{\varepsilon_i - n_{eff}^2} \right) = \nonumber\\
&& \qquad \qquad =
-\frac{\varepsilon_i}{\sqrt{|\varepsilon_o|\varepsilon_e}}
\sqrt{\frac{\varepsilon_e - n_{eff}^2}{\varepsilon_i - n_{eff}^2}},
\label{eq:ns1}
\end{eqnarray}
where a hyperbolic claddings with $\varepsilon_o < 0$ and
$\varepsilon_e > 0$ was taken into account and the definitions of
$p$, $q$ were used. As $hk_0 \geq 0$ the relation (\ref{eq:ns1}) has
solution only in following case
$$
n_{eff}^2 = \varepsilon_e, ~~ hk_0 = 0.
$$
But these conditions have no physical meaning (i.e., from this it
follows that $h=0$).

\section{Influence of Field Energy Density on Mode Propagation Characteristics}

This section is dedicated to analyze of the influence of propagating
mode energy on the mode propagation characteristics, such as an
effective refractive index and the cutoff frequencies in case
$\varepsilon_K < 0$.

In the previous sections a dimensionless parameter $\xi$ was used to
account nonlinear effects in all principal equations describing
guided wave. But $\xi$ has no clear physical meaning. So in the next
analysis we will consider a carried density of energy $W$ as a
parameter responsible for lever of nonlinearity influence.

The carried density of energy for planar wave in the dispersion-less
medium can be obtained from the Brillouin formula for this case and
takes a form:
\begin{equation} \label{eq:W}
W = \frac{1}{16 \pi} \int_{-\infty}^{\infty} \left( \varepsilon_e
|\tilde{E}_x|^2 +
    \varepsilon_o |\tilde{E}_z|^2  + |\tilde{H}_y|^2 \right) dx,
\end{equation}
where components of the TM wave field are defined by the equations
(\ref{eq:DEz}) and (\ref{eq:Hy}). The dielectric constants
$\varepsilon_e$, $\varepsilon_o$ are defined by (\ref{eq:epsl}).

\subsection{Effective Refractive Index of the TM Modes}

\begin{figure}[ht!]
    \centering
    \begin{minipage}[h]{.8\linewidth}
        \centering
        \includegraphics[width=\linewidth]{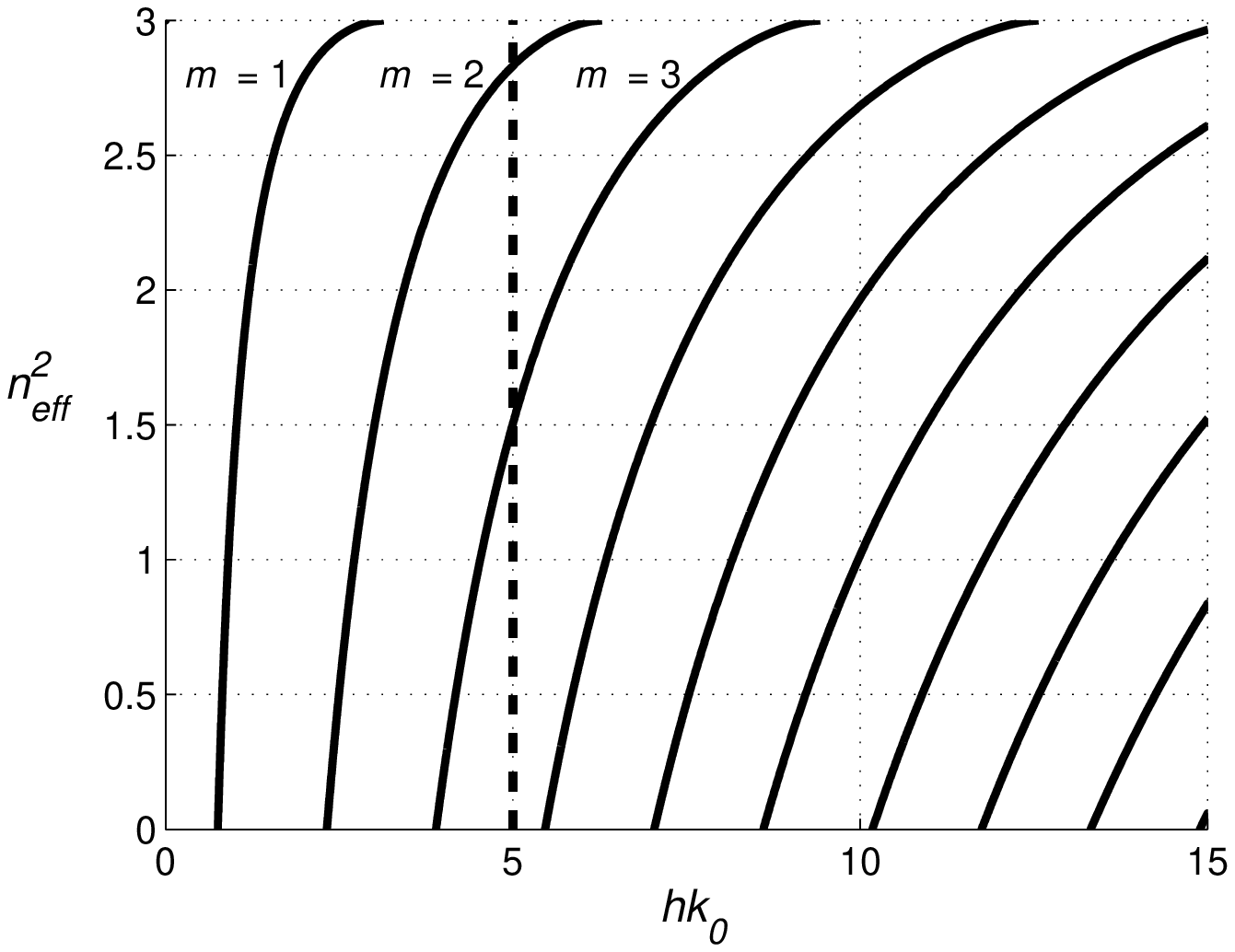}
        \\(\textit{a})
    \end{minipage}
    \vfill
    \begin{minipage}[h]{.8\linewidth}
        \centering
        \includegraphics[width=\linewidth]{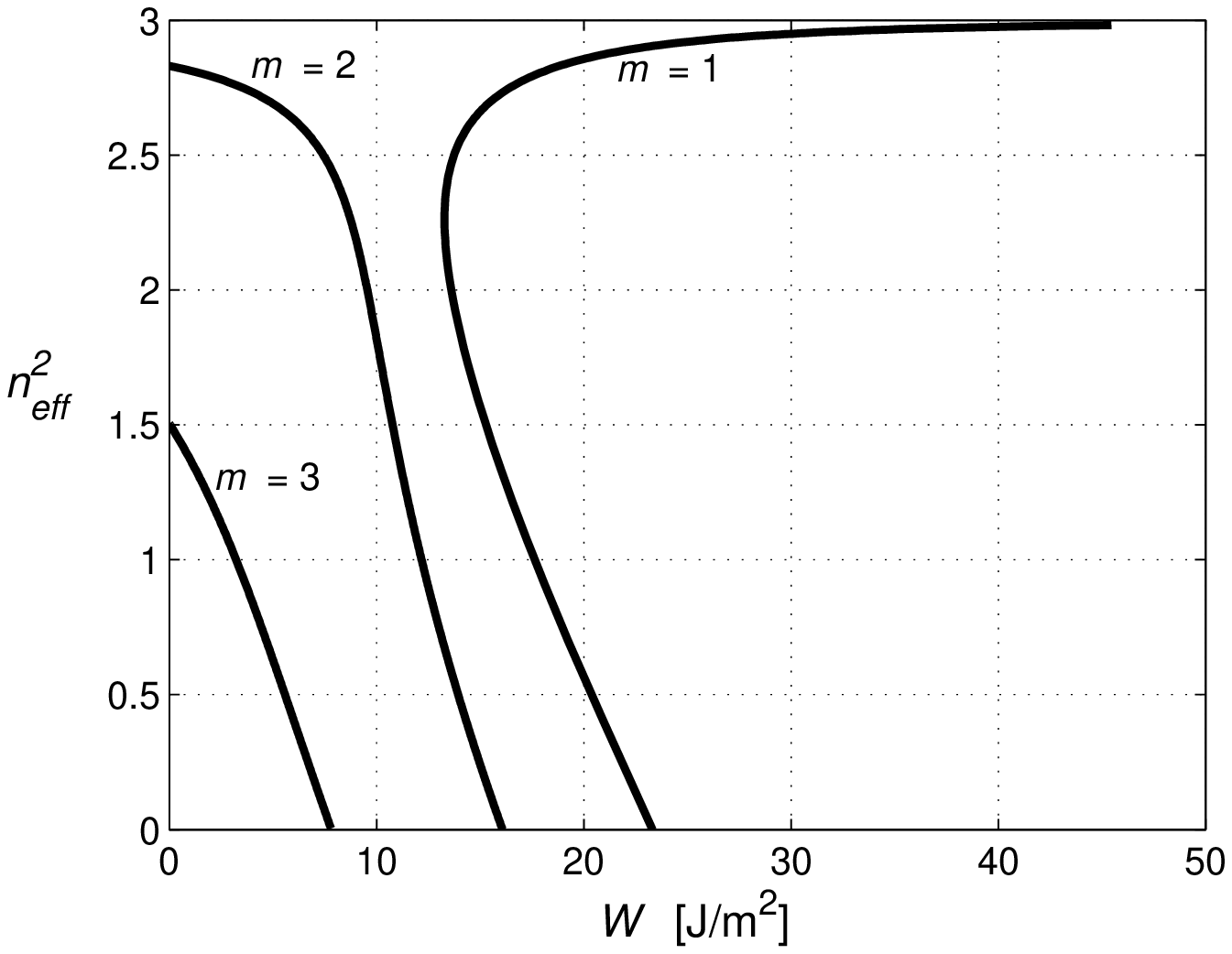}
        \\(\textit{b})
    \end{minipage}
    \vfill
    \begin{minipage}[h]{.8\linewidth}
        \centering
        \includegraphics[width=\linewidth]{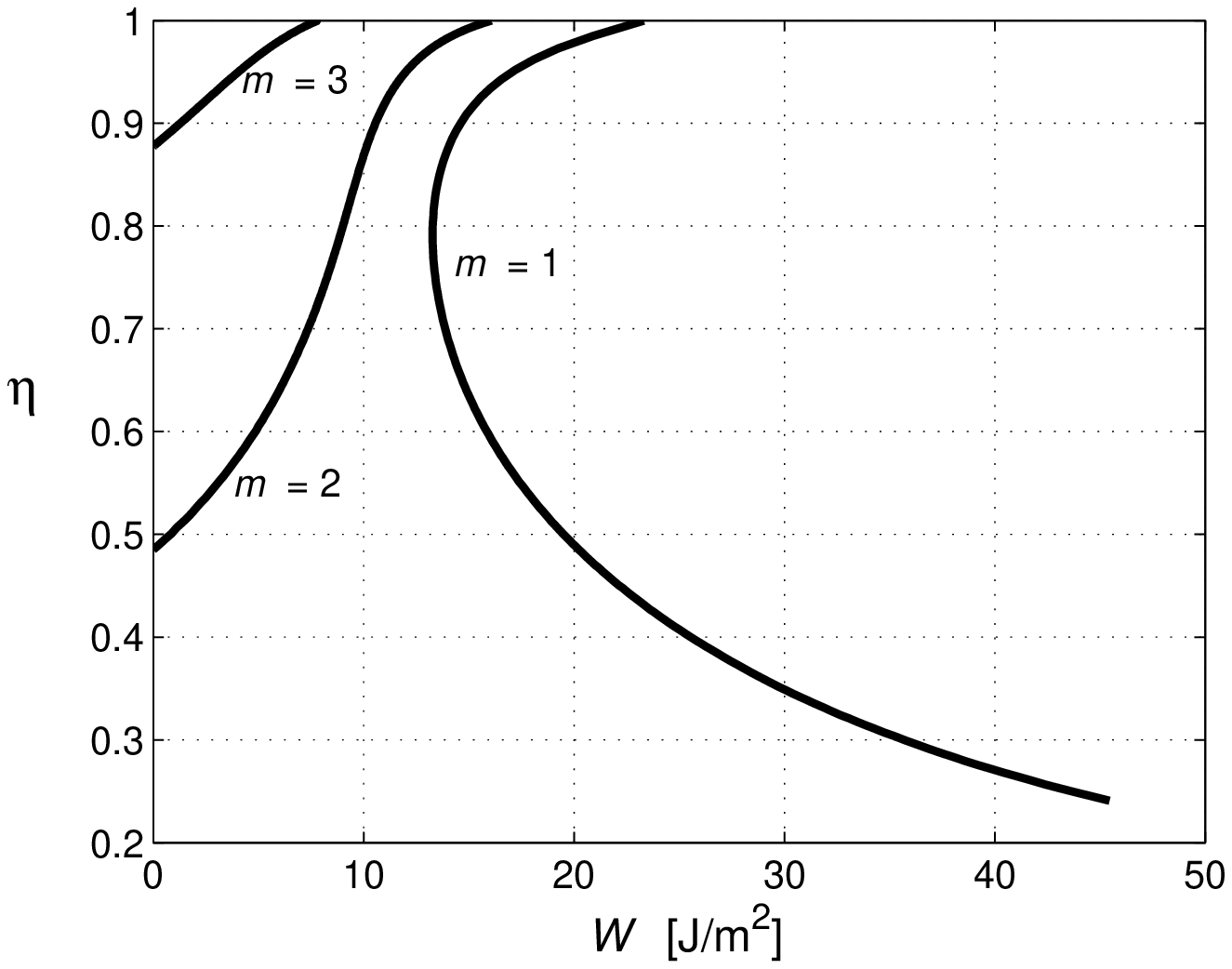}
        \\(\textit{c})
    \end{minipage}
    \caption{Hyperbolic waveguide with $hk_0 = 5$ ($a$) Dispersion curves for $\xi = 0$; ($b$) Dependence of effective refractive index on carried density of energy;
    ($c$) Dependence of energy fraction in the core on carried density of energy.}
    \label{Fig5:nefW}
\end{figure}

Let us consider a slab hyperbolic waveguide with fixed core width
and guided TM wave with constant frequency defined by equality $hk_0
= 5$. Values of the linear dielectric constant are the same as were
used in the previous section, $\varepsilon_K = -10^{-9}$ esu. In
Fig. \ref{Fig5:nefW} ($a$) dispersion curves for linear case are
presented. Dashed vertical line marks the value $hk_0 = 5$. From the
picture follows that in the linear case modes with marks $m=2$ and
$m=3$ can be exited in this waveguide. By comparing cases ($a$)
($b$) and ($c$) presented in the Fig.\ref{Fig3:disp:1} one can
notice that with increasing of the nonlinear parameter $\xi$ the
dispersion curves with marks $m=2$ and $m=3$ gradually leaves the
waveguide. Also at certain value of $\xi$ the mode with $m=1$ can be
exited in the waveguide while the modes with $m=2$ and $m=3$ can not
be exited.

An analogous behavior is illustrated in Fig. \ref{Fig5:nefW} ($b$).
The picture shows how the square of effective indexes of possible
modes varying with carried density of energy $W$. The core width and
radiation frequency satisfy condition $hk_0 = 5$. All possible in
this case mode marks are presented on the picture. At $W=0$ values
of $n_{eff}^2$ for marks $m=2$ and $m=3$ are the same as in the
linear case. They are also present in Fig. \ref{Fig5:nefW} ($a$) as
intersections of dashed line with corresponding dispersion curves.

With increasing of carried density of energy $W$ values of
$n_{eff}^2$ in cases $m=2$ and $m=3$ decrease more slowly at lower
values of $W$ and faster at higher values. That is because
dispersion curves presented in Fig.\ref{Fig5:nefW} ($a$) or
Fig.\ref{Fig3:disp:1}, Fig.\ref{Fig4:disp:2} are more flat at higher
$n_{eff}^2$ values. At certain value of $W$ mode with $m=3$ becomes
stopped, i.e., the associated effective mode index $n_{eff}^2$ or
propagation constant $\beta$ achieves zero value. After point
$n_{eff}^2 = 0$ guided mode becomes decaying and not propagating.
With further increasing of the electromagnetic field for mode with
$m=2$ the propagation constant also becomes equal to zero.

The mode with mark $m=1$ presents a special situation. It does not
exist in the linear case. But if $W$ exceeds a certain threshold
value mode with $m=1$ appears in the waveguide.
Another feature of this mode is that the function $n_{eff}^2 (W)$ is
double valued. One brunch of function has a similar behavior with
cases $m=2$ and $m=3$: $n_{eff}^2$ decreases to zero value. At
another brunch $n_{eff}^2$ tends to it maximum possible value
$n_{eff}^2 = \varepsilon_e$.

In Fig. \ref{Fig5:nefW} plot ($c$) shows the dependence of energy
fraction $\eta$ to whole value $W$. The fraction $\eta$ is defined
as
$$
\eta = \frac{W_{core}}{W},
$$
where $W_{core}$ is carried density of energy,which is concentrated
in the waveguide core only. Comparing the figures ($b$) and ($c$)
one can notice that with decreasing of $n_{eff}$ the guided wave
becomes more confined in the waveguide. This fact will be considered
in more details in the next subsection. If $n_{eff}^2$ tends to
$\varepsilon_e$ the considered mode leaves the waveguide core. Two
branches of the curve $m=1$ represent two guided modes. One of them
corresponds to mode that is confined in the core, another one
corresponds to mode with most of carrying energy concentrated in the
waveguide claddings.

\begin{figure}[h!]
    \centering
    \begin{minipage}[h]{.8\linewidth}
        \centering
        \includegraphics[width=\linewidth]{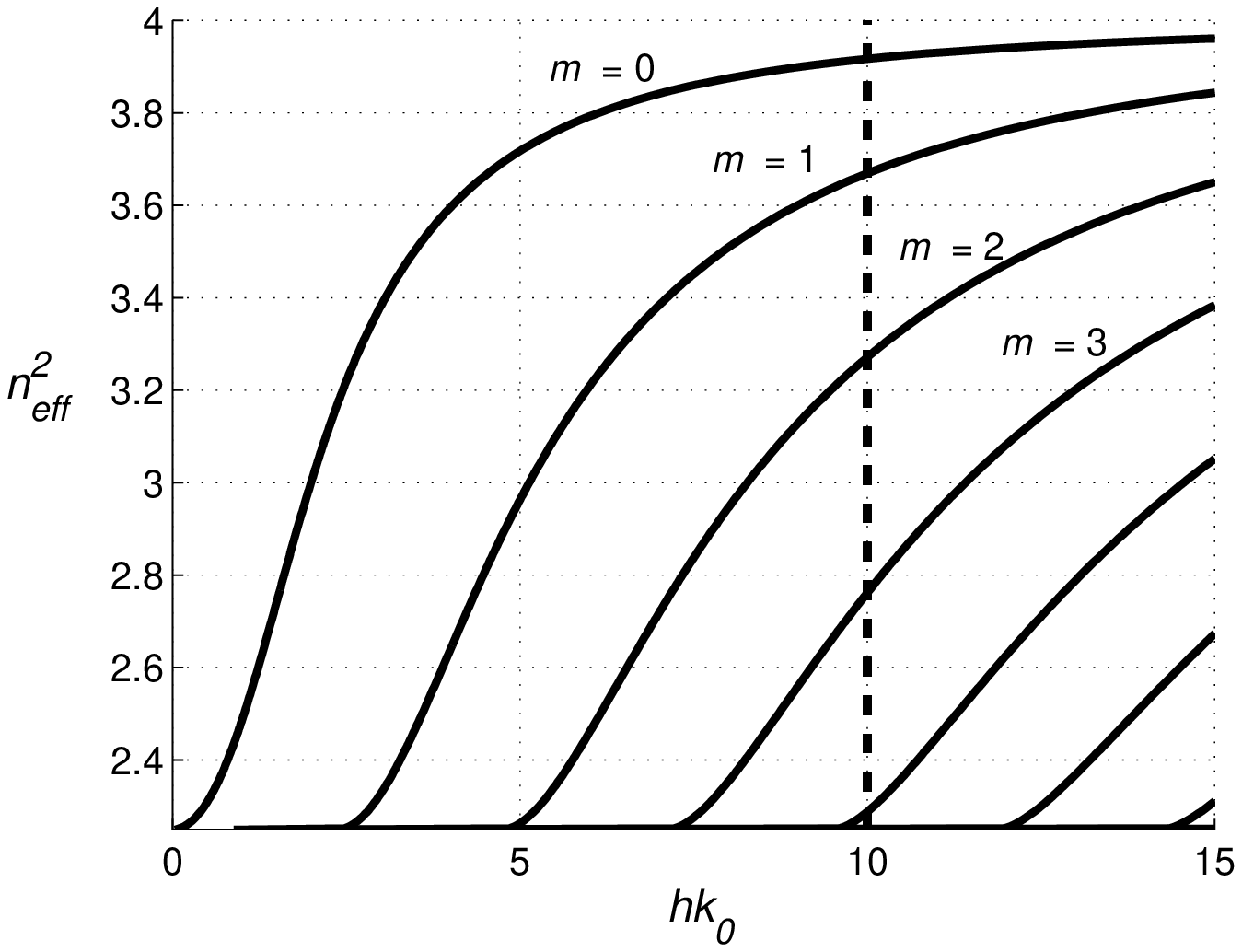}
        \\(\textit{a})
    \end{minipage}
    \vfill
    \begin{minipage}[h]{.8\linewidth}
        \centering
        \includegraphics[width=\linewidth]{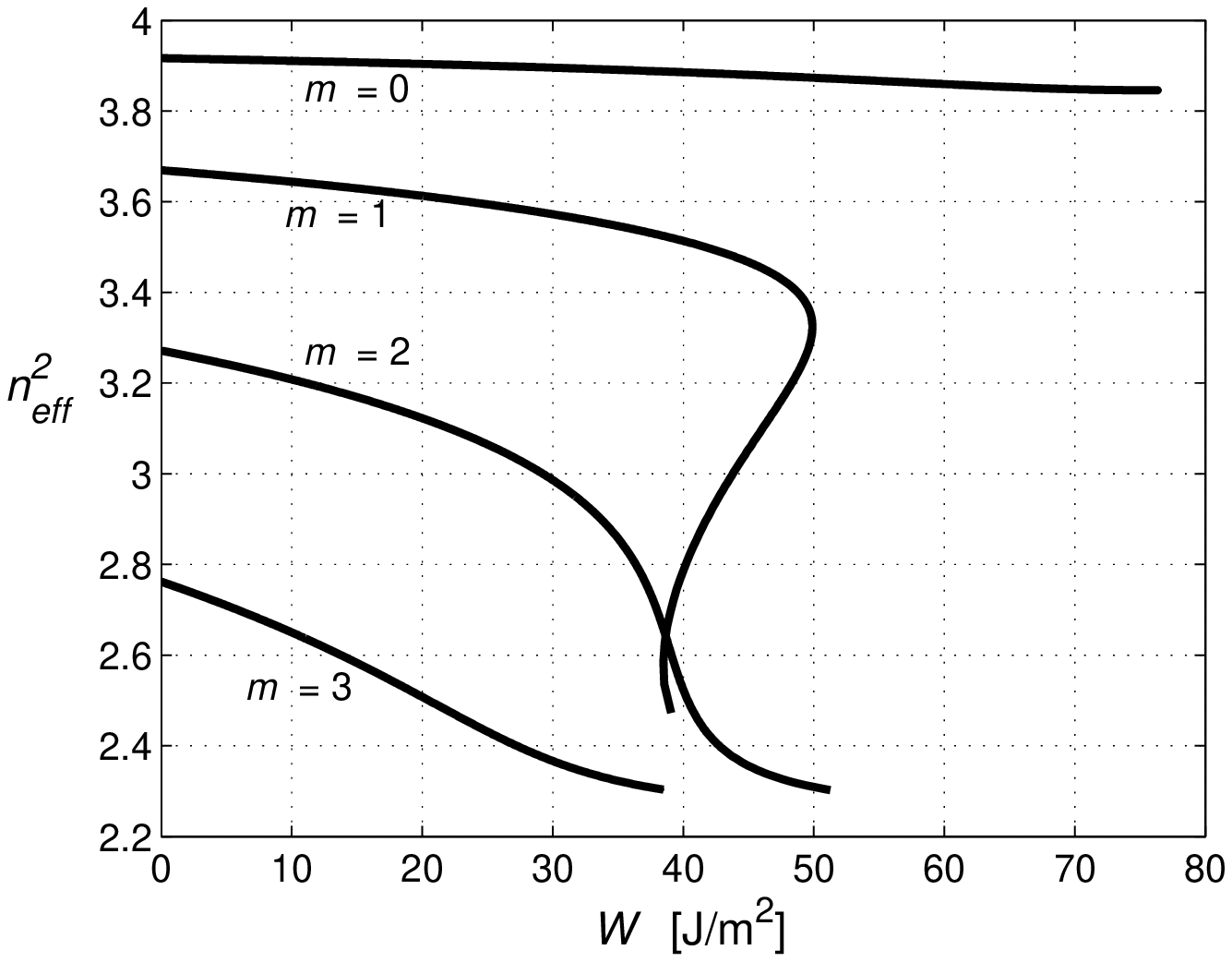}
        \\(\textit{b})
    \end{minipage}
    \caption{Usual dielectric waveguide with normalized core width $hk_0 = 10$. ($a$) Dispersion curves for $\xi = 0$;
    ($b$) Dependence of effective refractive index on carried density of energy.}
    \label{Fig6:nefWeli}
\end{figure}

For comparison with obtained results the similar results for the
usual dielectric slab waveguide with $\varepsilon_i = 4.0$,
$\varepsilon_e = \varepsilon_o = 2.25$, $\varepsilon_K = -10^{-9}$
esu are presented in Fig.\ref{Fig6:nefWeli}. Normalized core width
was proposed $hk_0 = 10$. In linear case modes with marks $m=0,
1,\ldots, 4$ can be exited. With increasing of the field intensity
the linear values of $n_{eff}^2$ for each modes are decreased
tending to theirs lowest values $n_{eff}^2 = \varepsilon_e =
\varepsilon_o$. No another curve appears. All possible mode are in
the linear case already.

\subsection{Decaying Length}

Due to (\ref{eq:DEz}) a parameter $k_{\vec{n}}=iq$ is a projection
of guided mode wave vector on normal direction to the waveguide
layers. Thus $q$ is a constant that defines how far wave intensity
can penetrate from the core to surrounding medium. Let us define a
decaying length as
$$
L_d = 1/2q,
$$
so at distance $L_d$ the field intensity $|E_z|^2$ decreases to
$1/e$ of its previous value. As $n_{eff}$ depends on field carried
density of energy $W$ the $L_d$ will also be changing with $W$.

\begin{figure}[ht!]
    \centering
    \begin{minipage}[h]{.8\linewidth}
        \centering
        \includegraphics[width=\linewidth]{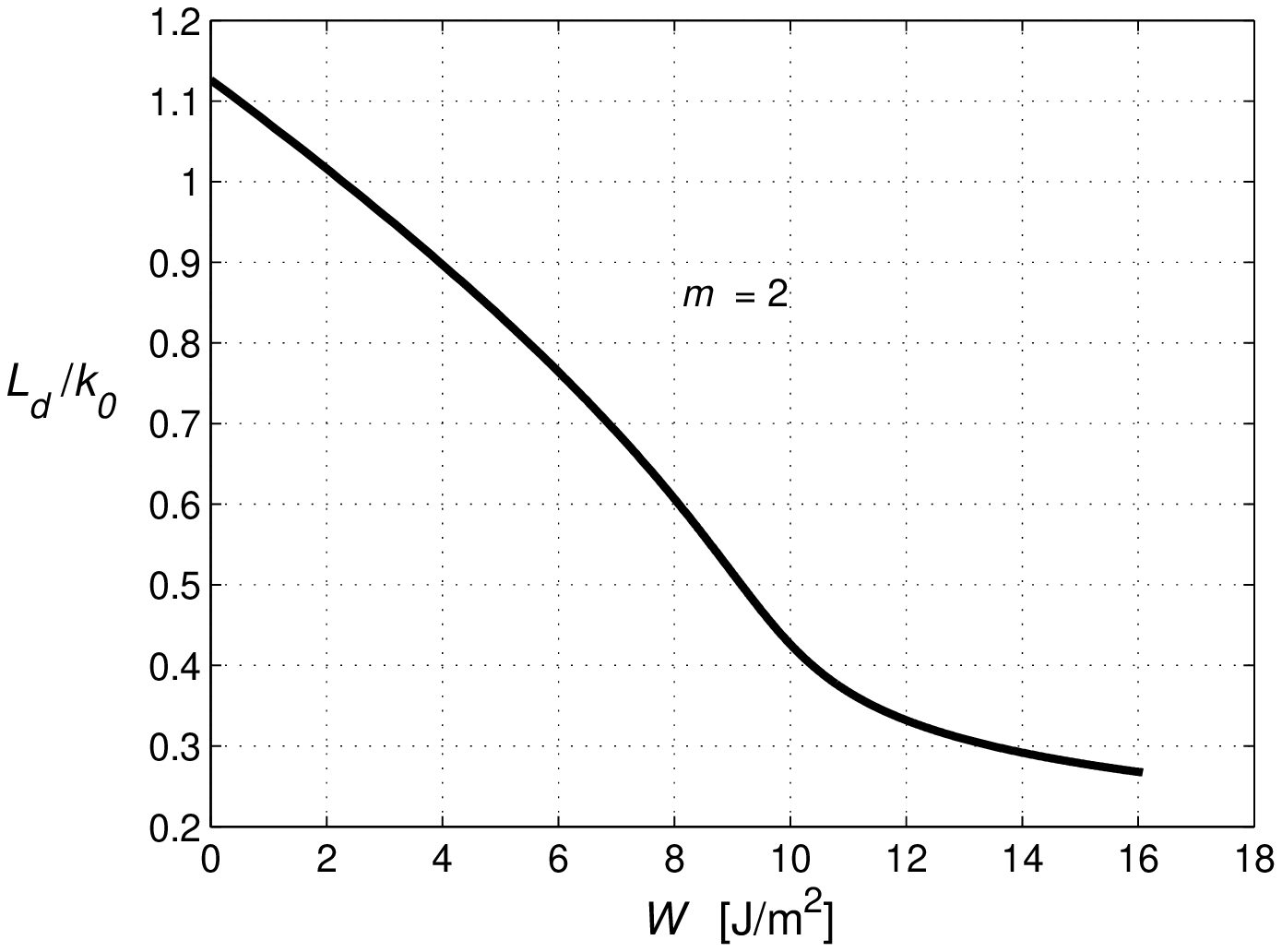}
        \\(\textit{a})
    \end{minipage}
    \vfill
    \begin{minipage}[h]{.8\linewidth}
        \centering
        \includegraphics[width=\linewidth]{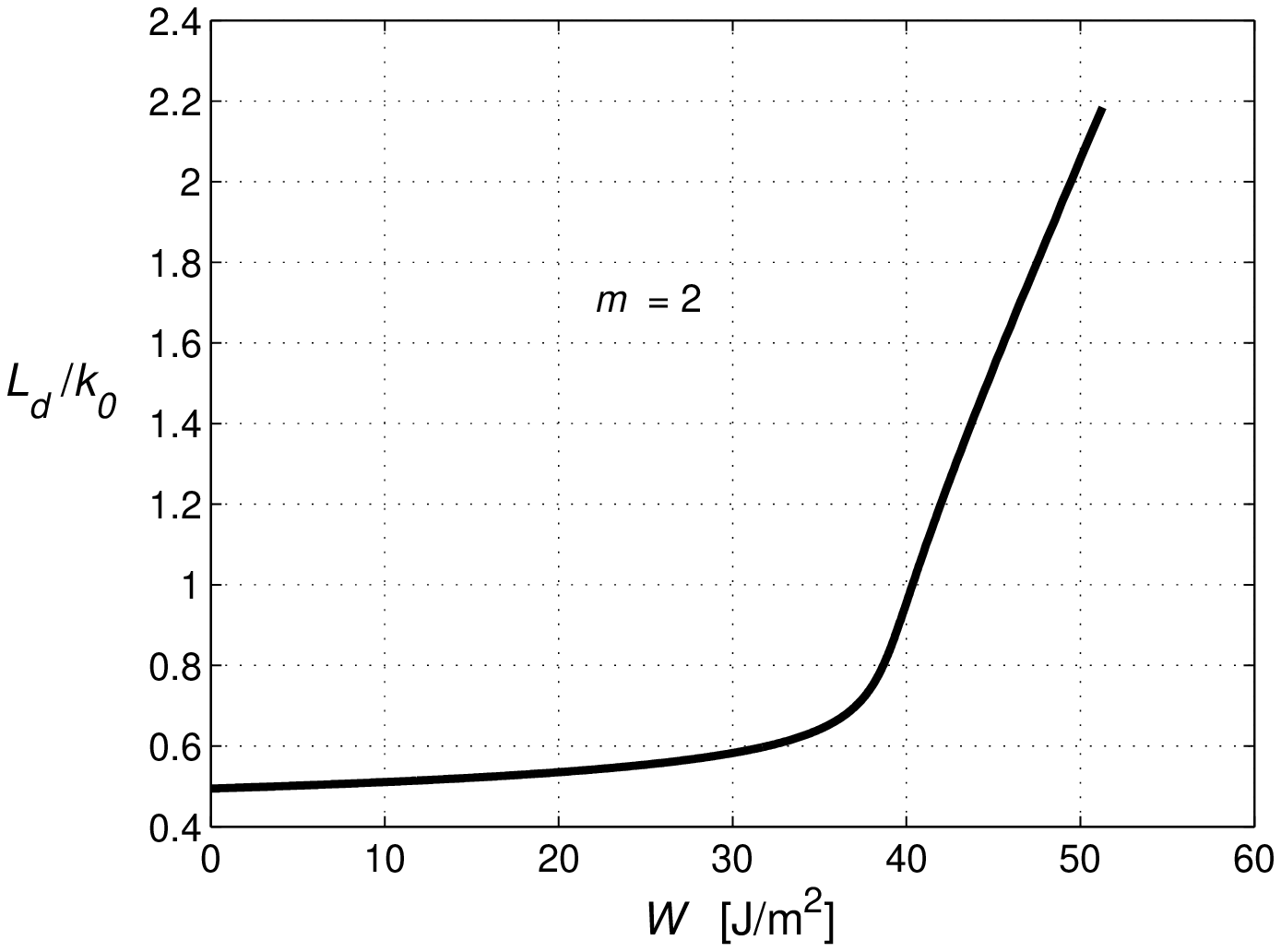}
        \\(\textit{b})
    \end{minipage}
    \caption{Dependence of decaying length in the claddings on
        surface density of energy. ($a$) Hyperbolic waveguide;
        ($b$) Usual dielectric waveguide.}
    \label{Fig7:dl}
\end{figure}

The Fig.\ref{Fig7:dl} illustrates this dependence both for a
hyperbolic waveguide ($a$) and a conventional waveguide case ($b$).
The mode with mark $m=2$ was chosen for both pictures.

As was shown earlier, in case of self-defocusing Kerr-medium of the
waveguide core the effective refractive index decreases with $W$. In
the hyperbolic case the value of $q$ increases with $n_{eff}$
decaying. Parameter $q$ is inversely proportional to decay length
$L_d$. Thus, the electromagnetic field will be more concentrated in
the waveguide core. Width of the considered
 mode becomes more narrow with $W$. In the usual dielectric waveguide case
at $\varepsilon_o>0$, $\varepsilon_e>0$ parameters $n_{eff}$ and $q$
are varying similarly. So with increasing of $W$ the electric field
penetrates farther into the waveguide cladding.

\begin{figure}[ht!]
\begin{minipage}{\linewidth}
    \begin{minipage}[h]{.49\linewidth}
        \centering
        \includegraphics[width=\linewidth]{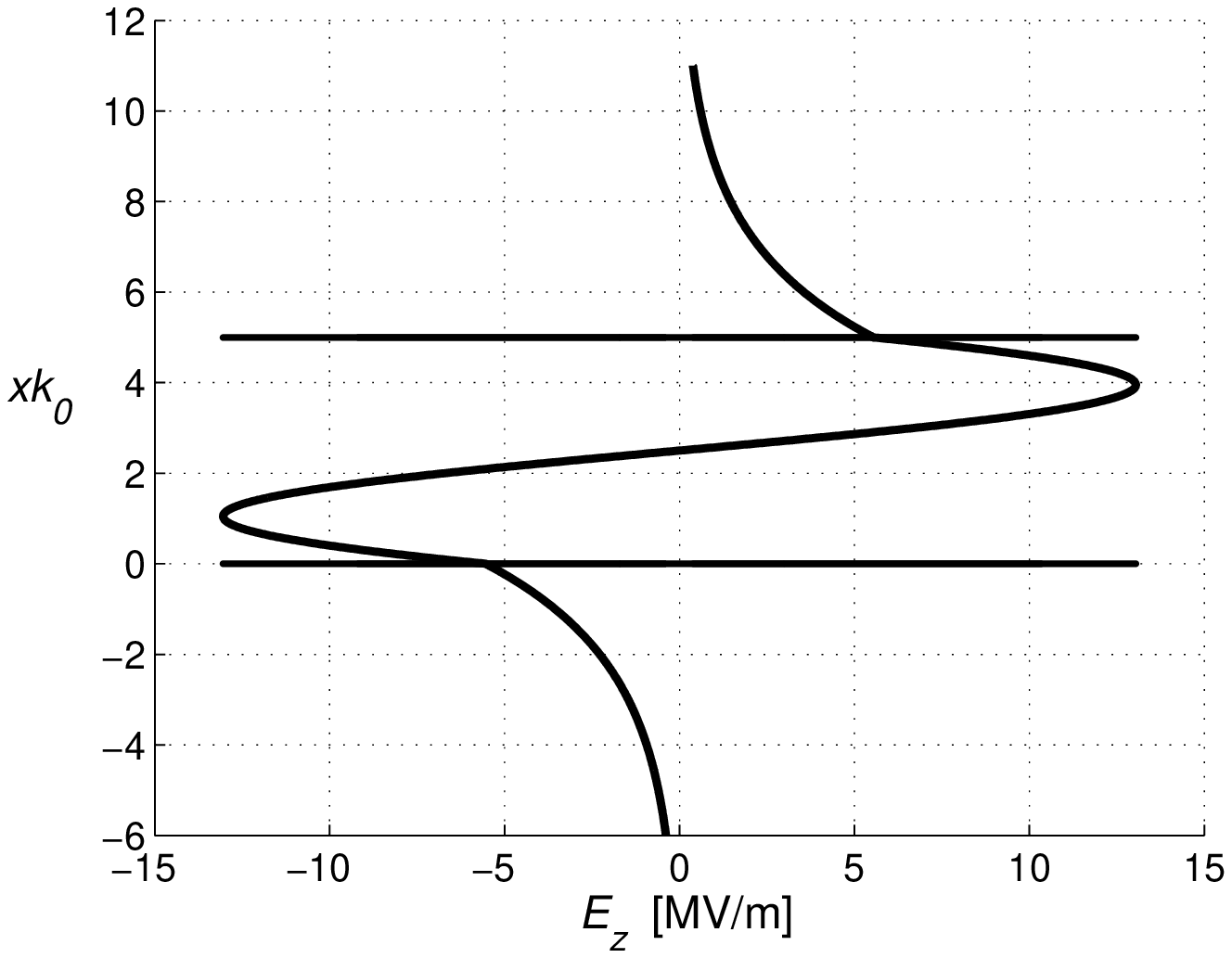}
        \\(\textit{a})
    \end{minipage}
    \hfill
    \begin{minipage}[h]{.49\linewidth}
        \centering
        \includegraphics[width=\linewidth]{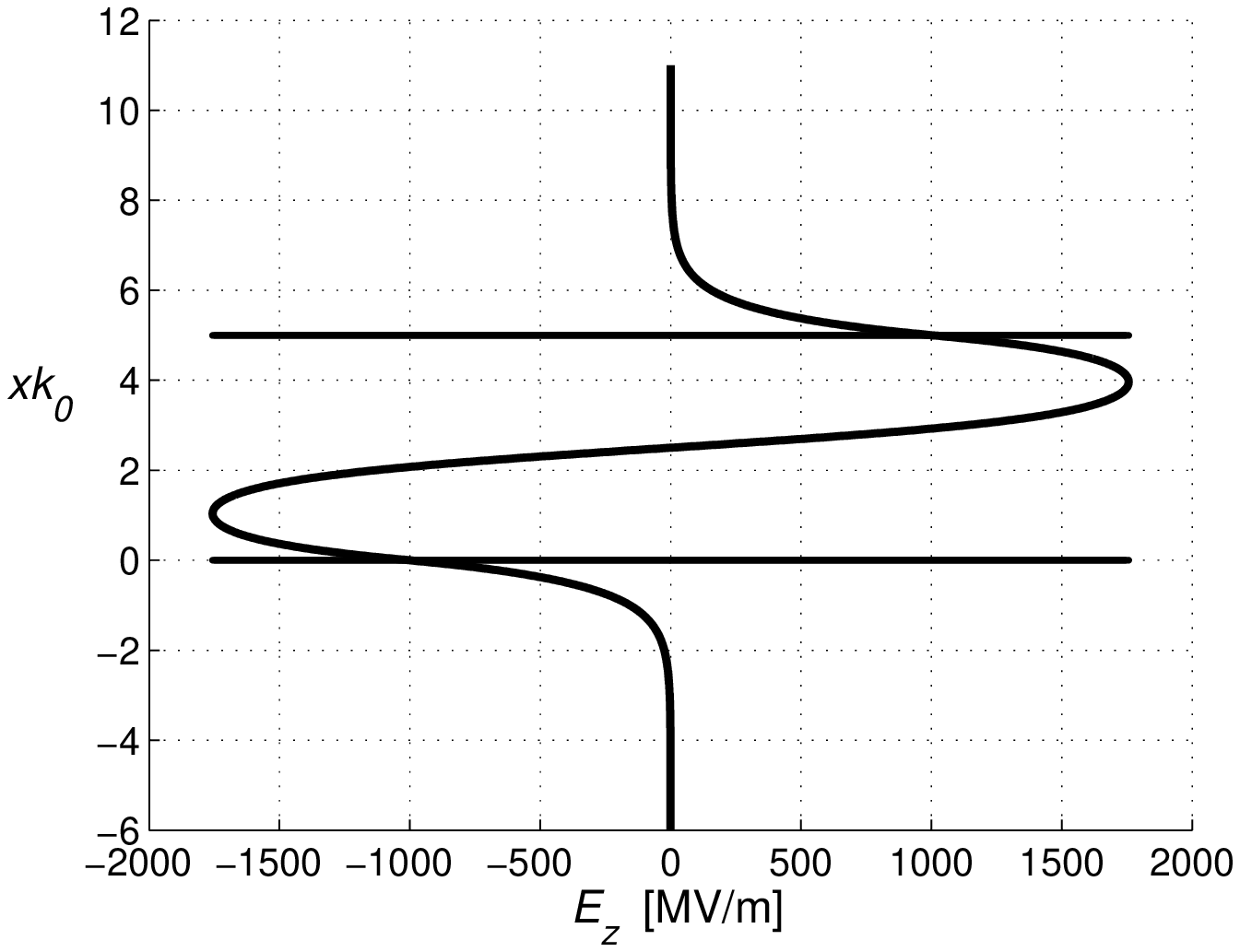}
        \\(\textit{b})
    \end{minipage}
\end{minipage}
%\vfill
    \begin{minipage}{\linewidth}
        \begin{minipage}[h]{.49\linewidth}
            \centering
            \includegraphics[width=\linewidth]{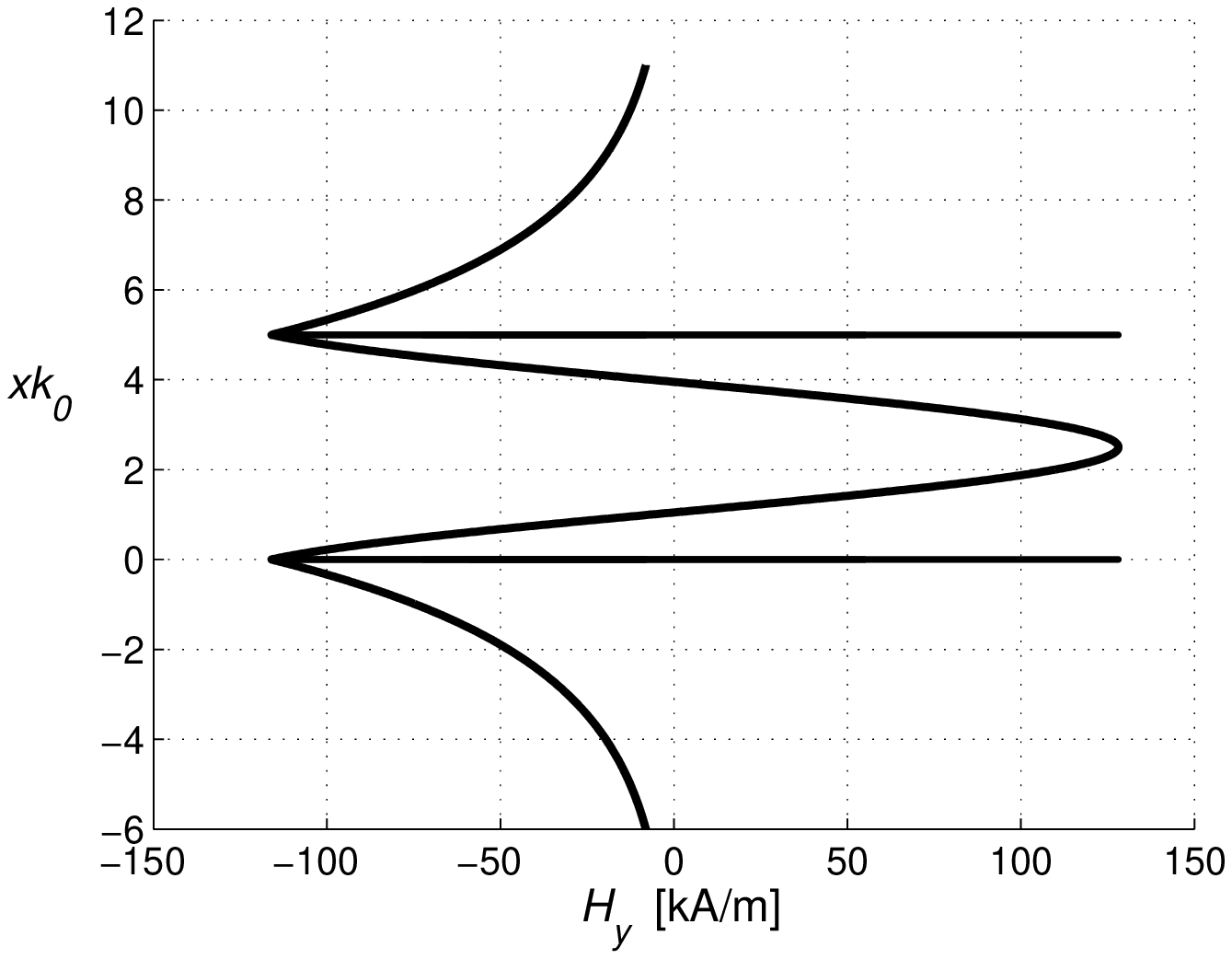}
            \\(\textit{c})
        \end{minipage}
        \hfill
        \begin{minipage}[h]{.49\linewidth}
            \centering
            \includegraphics[width=\linewidth]{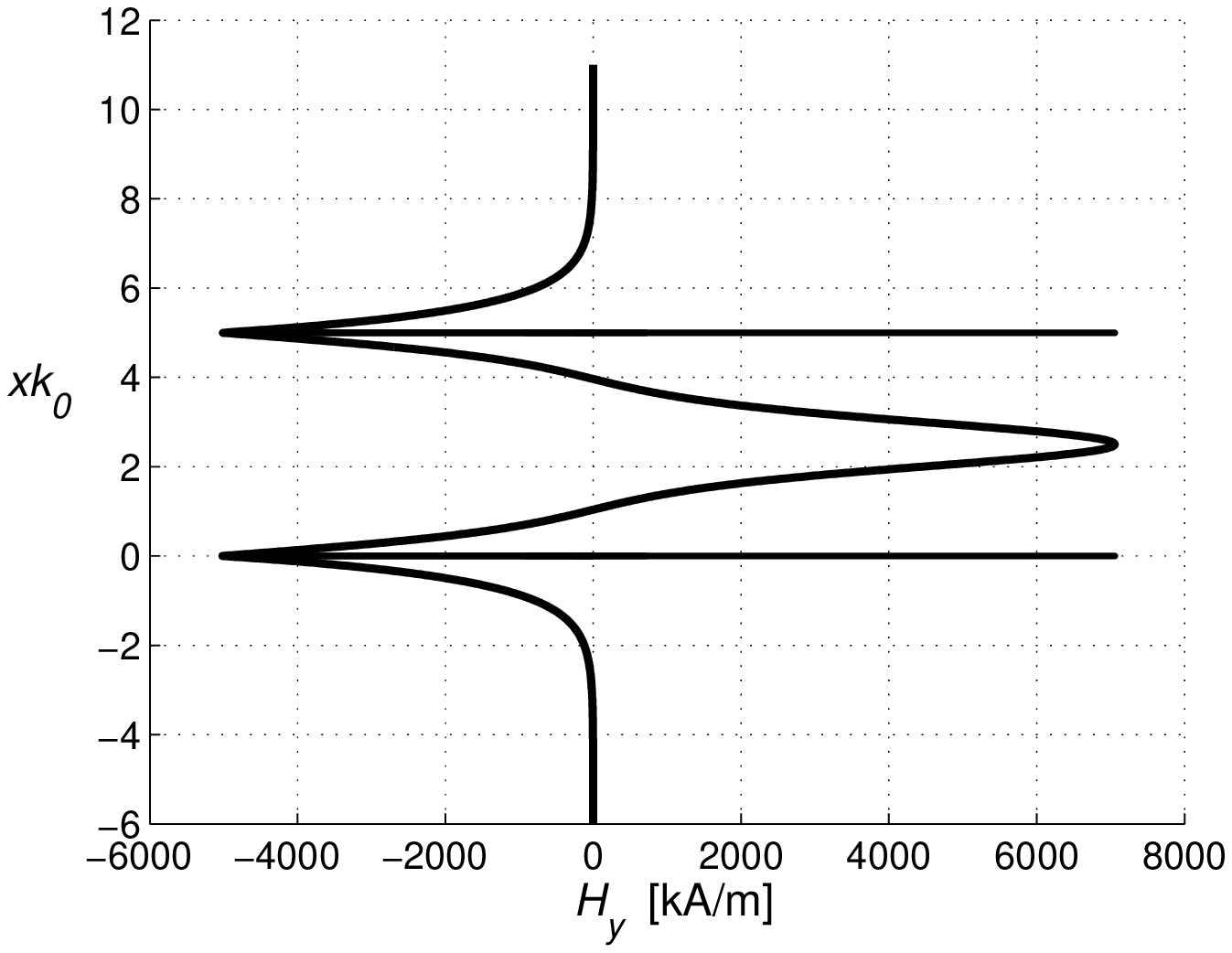}
            \\(\textit{d})
        \end{minipage}
    \end{minipage}
    \caption{Field distributions in the hyperbolic waveguide for TM mode $m=2$.
        ($a$), ($b$) Electric field distributions at low and high intensity correspondingly;
        ($c$), ($d$) Magnetic field distributions at low and high intensity.}
    \label{Fig8:fd}
\end{figure}

In Fig.\ref{Fig8:fd} the field distributions for mode $m=2$ in the
hyperbolic waveguide for lowest and highest values of $W$ are
presented. Normalized waveguide core width is $hk_0 = 5$. Cases
$(a)$ and $(c)$ correspond to $W \leq$ 1 J/m$^2$, $n_{eff}^2 =
2.28$. The cases $(b)$ and $(d)$ correspond to $W \geq 15$ J/m$^2$,
$n_{eff}^2 < 0.05$. The distributions of the electric fields outside
the core confirm the previous result presented in Fig.~\ref{Fig7:dl}
($a$).

\subsection{Cutoff Frequencies}

It was shown earlier that each TM mode of the hyperbolic waveguide
exists only in certain frequency range, i.e., there are two cutoff
frequencies at $\varepsilon_e < \varepsilon_i$. In this subsection
the influence of nonlinearity on cutoff frequencies of mode with
index $m=2$ will be tested.

\begin{figure}[h!]
    \centering
    \begin{minipage}[h]{.8\linewidth}
        \centering
        \includegraphics[width=\linewidth]{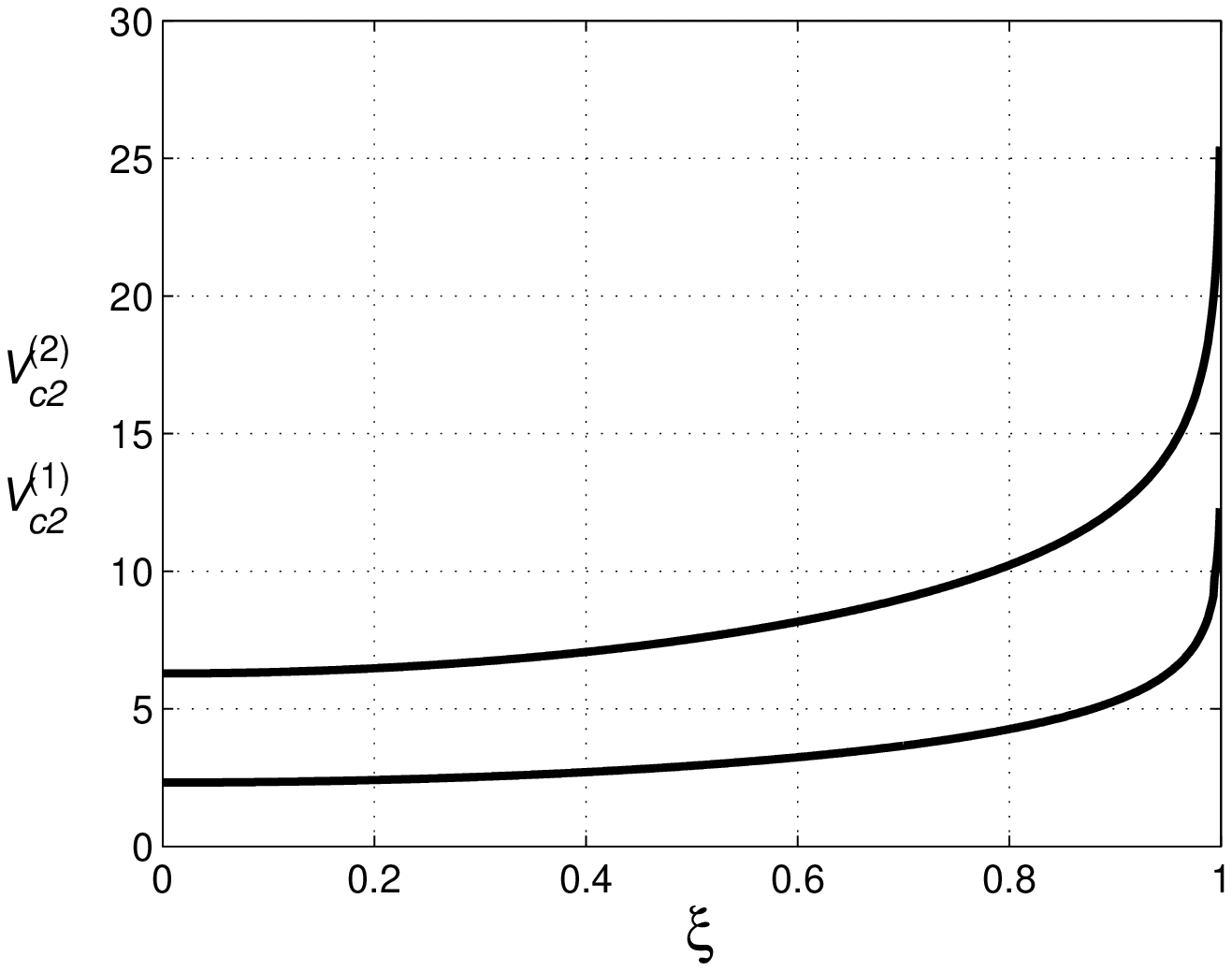}
        \\(\textit{a})
    \end{minipage}
    \vfill
    \begin{minipage}[h]{.8\linewidth}
        \centering
        \includegraphics[width=\linewidth]{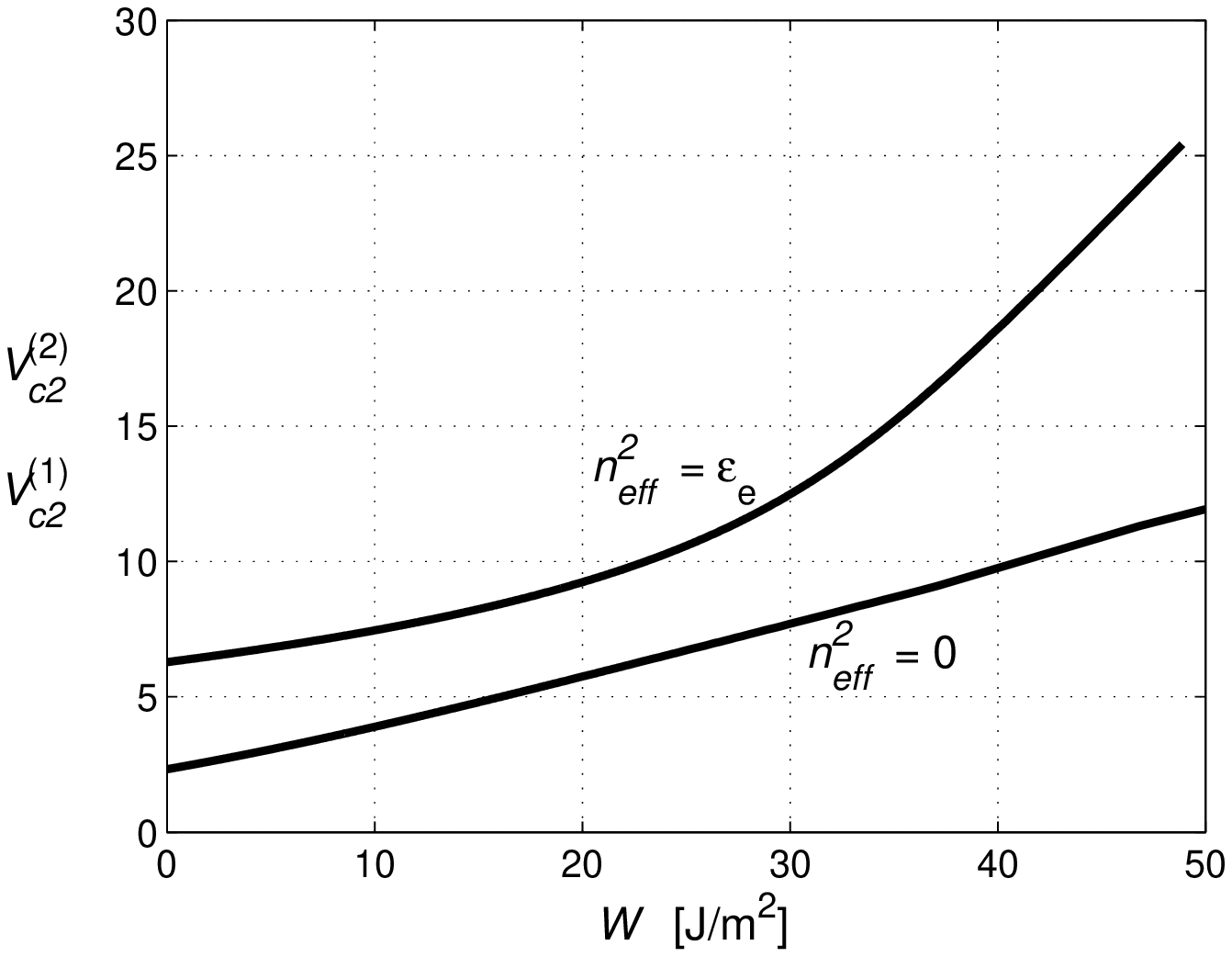}
        \\(\textit{b})
    \end{minipage}
    \caption{Dependence of cutoff frequencies on  $\xi$ ($a$) and  $W$ ($b$).}
    \label{Fig9:cut}
\end{figure}

Let us define mode cutoff frequencies as
\begin{equation}\label{eq:f8a}
    V_{cm}^{(1)} = hk_0(n_{eff}^2 = 0),~~~V_{cm}^{(2)} = hk_0(n_{eff}^2 = \varepsilon_e),
\end{equation}
where $m$ is a mode mark ($m = 1,2,\ldots $) and the normalized
frequency (or normalized width) $hk_0$ is solution of
(\ref{eq:disp2}). So a slab hyperbolic waveguide with core width $h$
confines only TM mode for radiation frequencies lying in the
corresponding interval
\begin{equation} \label{eq:fre}
\frac{c}{h} V_{cm}^{(1)} \leq \omega \leq \frac{c}{h} V_{cm}^{(2)}.
\end{equation}
Cutoff frequencies could be obtained from the equation
(\ref{eq:disp2}) as functions of $\xi$ at appropriate values of
parameters $p(n_{eff}^2)$ and $q(n_{eff}^2)$. The results for $m=2$
are presented in Fig.\ref{Fig9:cut} ($a$). Initial values of
$V_{c2}^{(1)}$ and $V_{c2}^{(2)}$ at $\xi=0$ were defined from the
linear dispersion equation obtained in \cite{Lyashko:15}. The
influence of field carried density of energy on the cutoff
frequencies are presented in Fig.\ref{Fig9:cut} ($b$). Dielectric
constants are the same as in previous sections. Kerr constant was
chosen to be equal to $\varepsilon_K = -10^{-9}$ esu.

From the picture it follows that both cutoff frequencies increase
with $\xi$ (intensity). The width of interval ($V_{cm}^{(1)}$,
$V_{cm}^{(2)}$) slowly increase with $\xi$.  At values $\xi \to 1$
both frequencies tend to infinity, that means that there are no
guided waves in this waveguide.

Strictly speaking at limit value $\xi = 1$ the following results for
cutoff frequencies can be obtained
$$V_{cm}^{(1)} \to \infty,\quad V_{cm}^{(2)} = 0.$$
Thus the function $V_{cm}^{(2)} (\xi)$ is continuous function in
interval $[0, 1)$.

\section{Conclusion}

Properties of the TM guided modes in the hyperbolic slab waveguide
are theoretically investigated. The slab waveguide is formed from a
non-dissipative dielectric core and a hyperbolic environment with
anisotropy axis aligned with normal vector to the layers interfaces.
The nonlinear Kerr response of the core was taken into account. Only
evanescent waves are in the hyperbolic environment. So possible
absorption of the hyperbolic medium was not considered. The effect
of the dissipation has been discussed in \cite{LM:16}. As usually
the losses lead to the guided wave amplitude decreasing. However, it
was fond that if the carry wave frequency lies in transparency
region of the hyperbolic medium the principal phenomena presented in
the paper will exist and the effects of losses will be like effects
in the case of waveguide with metallic covering layer.

The dispersion relation for the guided modes was found. It was shown
that if dielectric constant of the core is greater then
extraordinary permittivity of the environment each TM guided wave
has two cutoff frequencies as well as in works \cite{Lyashko:15,
LM:16}. So each mode can be exited only in appropriate frequency
range. Contrary to the case considered in \cite{LM:16}, here the
principal part of the guided wave energy in the waveguide with
nonlinear core is localized in nonlinear dielectric medium. In the
case of self-focusing Kerr medium of the core a number of possible
modes increases with the field intensity. In case of self-defocusing
medium this dependence is inverse.

The value of mode propagating constant decreases with intensity of
the radiation field in the self-defocusing case. Hence the
propagations constant will be zero at certain value of the
intensity. With further increasing of field intensity mode becomes
an evanescent and disappears. Thus varying the energy of the field
one can obtain the slow-light waveguide of a new kind. The changing
of a number of possible waveguide modes in the hyperbolic waveguide
under consideration could be exploited to fabrication of the
all-optical modulator or another all-optical devices.

The influence of carried density of energy on the mode width and
cutoff frequencies was also analyzed in the self-defocusing core
medium case. It was shown that with field energy growth mode width
becomes more narrow as opposite to the usual dielectric waveguide
case.

\section*{ Acknowledgement}
%\ack

We are grateful to Prof. I. Gabitov and Dr. C. Bayun for
enlightening discussions. This investigation is funded by Russian
Science Foundation (project 14-22-00098).

%\References

%\section*{ References}

%\include{referense2}

\end{document}